\documentclass{ws-rv9x6}
\begin{document}

\setcounter{chapter}{3}
\chapter{PHASE TRANSITIONS IN FRUSTRATED \\
VECTOR SPIN SYSTEMS: NUMERICAL STUDIES}
\markboth{D. Loison}{Frustrated vector spin systems: numerical
studies}

\author{Damien Loison}
\addcontentsline{toc}{author}{D. Loison}
\address{
Institut f\"ur Theoretische Physik, Freie Universit\"at Berlin,
Arnimallee 14, 14195 Berlin, Germany\\
Damien.Loison@physik.fu-berlin.de }

\setcounter{footnote}{0}
\setcounter{figure}{0}
\setcounter{equation}{0}
\setcounter{section}{0}

\section*{Abbreviations}
Some abbreviations are used in this chapter (by alphabetic order):\\
bct=body-centered-tetragonal, BS=Breakdown of Symmetry,
$d$=dimension, fcc=face-centered cubic lattice, FSS=Finite Size
Scaling, GS=Ground State, hcp=hexagonal-close-packed lattice,
KT=Kosterlitz-Thouless, $L$=system size, MC=Monte Carlo,
MCRG=Monte Carlo Renormalization Group, $N$= number of components
of the spin {\bf S}, NN=Nearest Neighbors, NNN=Next Nearest
Neighbors, RG=Renormalization Group, STA=Stacked Triangular
Antiferromagnetic lattices, STAR=Stacked Triangular
Antiferromagnetic lattices with Rigidity, $T$=Temperature,
$T_c$=critical Temperature, $V_{N,P}$=Stiefel model.

\section{Introduction}
We present in this chapter a review on recent numerical studies
dealing with frustrated vector spin systems in two and three
dimensions. A system of spins is frustrated\index{frustration,
definition} when all interactions between spin pairs cannot have
simultaneously their optimal values. In other words a system is
frustrated when the global order is incompatible with the local
one, a definition applicable in a broader sense and not restricted
to spins. For spin systems frustration has the consequence that
the ordered state is different from the collinear order found for
common unfrustrated antiferromagnets or ferromagnets. To study and
to classify the phase transitions between the ordered and less
ordered states the symmetries these states play an important role.
The critical behavior of second order transitions is principally
governed by the change of symmetries. Therefore even very
different systems like superfluid helium and XY spin systems can
have the same critical properties. This fact is called
universality and our objective is to analyze the corresponding
universality classes for frustrated spin systems. We will mainly
review magnetic models since they are the easiest to analyze
numerically and theoretically. Nevertheless the results should be
valid for any system belonging to the same universality class.

During the last decade important progress has been made in the
understanding of the physics of frustrated spin systems. For example
there is now convincing evidence on the genuine first order nature
of the phase transition for $XY$ and Heisenberg spins for the three
dimensional Stacked Triangular Antiferromagnet (STA). This still contradicts
the latest renormalization group expansion based on resummation.
We think that studying phase transitions of vector spins theoretically,
numerically and experimentally should have implications beyond this
special field, that is for the understanding and the theory of phase
transition in general.

We will concentrate our attention to the phase transition of the
physical $XY$ and Heisenberg spin systems in two and three
dimensions. However, to understand these systems we have to
analyze also the phase transitions of frustrated spins of $N$
components, where $N$ takes all integer values from 1 to $\infty$
and not only 2 and 3 for $XY$ and Heisenberg spins. We will also
present studies for dimension $d$ varying between two and four.
Then we will review the particular case of strictly two dimensions
where topological defects have a dominant role.

Frustrated Ising models are reviewed also in this book by Diep and
Giacomini (chapter 1) and by Nagai, Horiguchi and Miyashita
(chapter 2).

Since most of numerical simulations presented here use the Monte
Carlo method, a short appendix (\ref{appendixMC}) at the end is
devoted to this technique. In addition, since the renormalization
group is fundamental to the understanding of phase transitions, a
small appendix (\ref{appendixRG}) is added to discuss the methods
used here for frustrated spin systems.

\section{Breakdown of Symmetry\index{breakdown of symmetry}}

We first briefly review the fundamental concept of the reduction
or the Breakdown of Symmetry (BS) in the transition from the high-
to the low-temperature phases. The classification of phase
transitions in universality classes\index{universality class}  is
based on this concept.

\subsection{Symmetry in the high-temperature region}
We will consider the Hamiltonian:
\begin{eqnarray}
\label{eq_H_ON}
H = -J_1 \sum_{(ij)} {\bf S}_{i}.{\bf S}_{j}
\end{eqnarray}
where ${\bf S}_{i}$ are  $N$ component classical vectors of unit
length and the sum is usually restricted to the nearest neighbors
or at least to short range interactions. The symmetry of this
Hamiltonian is $O(N)$ or equivalently $Z_2 \otimes SO(N)$ where
$O(N)$ is the orthogonal transformation group of $N$-dimensional
Euclidean space. For $SO(N)$ the determinant is unity, and the
Ising symmetry $Z_2$ corresponds to the mirror operation. If we
add other terms, like dipolar interactions or anisotropies, the
symmetry group will be reduced. Some other changes, like a cubic
term $({\bf S}_{i}.{\bf S}_{j})^3$, will not reduce the symmetry.
In many experiments anisotropies will reduce Heisenberg ($N=3$)
symmetry to $XY$ ($N=2$) or Ising ($N=1$) symmetry. Long-range
interactions could also be present and the interpretation of
experimental results could be problematic (see the crossover
section in \ref{appendixRG}).

We will also encounter the Potts\index{Potts model}  symmetry
$Z_q$. The Potts model has the Hamiltonian:
\begin{eqnarray}
\label{eq_H_Potts}
H = -J_1 \sum_{(ij)} \delta_{q_iq_j}
\end{eqnarray}
$\delta_{q_iq_j}$  refers to the $q$ states Potts spin with
$\delta_{q_iq_j}=0$ when $q_i\ne q_j$ and
$\delta_{q_iq_j}=1$ when $q_i = q_j$.

In addition we have to consider the symmetry of the lattice which is the
sum of all symmetry elements which let the lattice invariant.
For example, the triangular lattice has $C_{3v}$ symmetry
and the square lattice $C_4$ symmetry.\cite{Inui_group}
Therefore we have to take into account the total BS composed of the
$O(N)$ spin rotations and the symmetries of the lattice.

\subsection{Breakdown of symmetry for
ferromagnetic systems}\index{breakdown of symmetry}
\index{breakdown of symmetry, ferromagnets} If $J_1$ in
(\ref{eq_H_ON}) is taken positive the Ground State (GS), for any
lattice, is ferromagnetic with collinear spins pointing in the
same direction. If $J_1$ is negative (antiferromagnetic) but there
is no frustration like for the square lattice, the GS is collinear
with alternating direction. Then the BS will be identical to the
ferromagnetic case. The symmetry at low temperature is $O(N-1)$
for the spins and the lattice symmetries are preserved. Therefore
the BS  is $O(N) \longrightarrow O(N-1)$ where $O(N)$ is the
symmetry of the phase at high temperatures and $O(N-1)$ the
symmetry at low temperatures. We will denote this BS
$O(N)/O(N-1)$. Keeping in mind that $O(0)\equiv 1$, $O(1) \equiv
Z_2$, $SO(1)\equiv 1$ and $O(N) \equiv Z_2 \otimes SO(N)$ for
$N\geq 2$, the BS $O(N)/O(N-1)$ can be written as $Z_2$ for $N=1$,
$SO(2)$ for $N=2$, and $SO(N)/SO(N-1)$ for $N>2$.

All the ferromagnetic systems (cubic, stacked triangular, $\dots$)
with a Hamiltonian of type (\ref{eq_H_ON}) have an identical
BS and consequently belong to the same universality class.
The only relevant variables
are the dimension of the spin space, i.e. the number of components of
the spin $N$, and the dimension of the
real space $d$.

For Potts\index{Potts model}  ferromagnetic system with a
Hamiltonian of the type (\ref{eq_H_Potts}) and $J_1>0$, the Potts
symmetry is broken in the low temperatures phase and the BS is
simply $Z_q$. Again this does not depend on details of the system.
This model has a second order phase transition for $q\leq q_c$ and
a first order phase transition for $q>q_c$. We know that $q_c=4$
in two dimensions and $q_c<3$ in three dimensions. Therefore, in
three dimensions, for any $q\geq 3$ the transition will be of
first order.

In Table~\ref{table_ferro} we give the value of the critical
exponents calculated by the Renormalization Group\cite{Antonenko1}
(RG) depending on $N$ and $q$ in three dimensions ($d=3$). Even if
$N>3$ does not correspond to physical systems, we will get very
useful information for frustrated systems.

\begin{table}
\tbl{ Critical\index{critical exponents} exponents for the
ferromagnetic systems in three dimensions calculated by RG.
$^{(a)}$We cannot define exponents in a first-order transition,
however in the case of a weak first-order transition the exponents
found with MC and in experiments must tend to these values.
$^{(b)}$calculated by $\gamma/\nu=2-\eta$. \label{table_ferro} }
{\tabcolsep10pt
\begin{tabular}{@{}c|@{\hspace*{0.1cm}}c|@{\hspace*{0.1cm}}c|@{\hspace*{0.1cm}}c|@{\hspace*{0.1cm}}c|@{\hspace*{0.1cm}}c}
\hline\\[-3pt]
{\bf BS} &{\bf $\alpha$} &{\bf $\beta$} &{\bf $\gamma$} &{\bf $\nu$} &{\bf $\eta$} \\[2pt]
\hline\\[-8pt]
$O(1)/O(0)\equiv Z_2$&0.107&0.327&1.239&0.631&0.038\\[2pt]
\hline\\[-8pt]
$O(2)/O(1)\equiv SO(2)$&-0.010&0.348&1.315&0.670&0.039\\[2pt]
\hline\\[-8pt]
$O(3)/O(2)\equiv SO(3)/SO(2)$&-0.117&0.366&1.386&0.706&0.038\\[2pt]
\hline\\[-8pt]
$O(4)/O(3)\equiv SO(4)/SO(3)$&-0.213&0.382&1.449&0.738&0.036\\[2pt]
\hline\\[-8pt]
$O(6)/O(5)\equiv SO(6)/SO(5)$&-0.370&0.407&1.556&0.790&0.031\\[2pt]
\hline\\[-8pt]
1$st$ order$^{(a)}$ ($Z_q$ if $q>3$)&1&0&1&1/3&-1$^{(b)}$\\[2pt]
\hline
\end{tabular}}
\end{table}

\subsection{Breakdown of symmetry
for frustrated systems}\index{breakdown of symmetry, frustrated
systems} The BS in frustrated systems is more complicated. We will
present examples where the $O(N)$ symmetry is reduced to $O(N-1)$,
$O(N-2)$, $\cdots$, $O(O)$ (completely broken in the last case).
In addition it is possible that the lattice symmetry is also
reduced, usually giving an additional $Z_q$ (mostly $q=2$ and
$q=3$) broken symmetry. Therefore the possible BS are $S_{lattice}
\otimes O(N)/O(N-P)$ with $S_{lattice}=1$, $Z_2$ or $Z_3$ and $P$
varies from 1 to $N$.

\subsubsection{Stacked\index{STA}
triangular antiferromagnetic lattices
\label{STA_III}}\index{stacked triangular antiferromagnets, STA}

 We consider here the stacked triangular
antiferromagnetic lattice (STA) with  nearest-neighbor (NN) and
next-nearest-neighbor (NNN)  $J_2$ antiferromagnetic interactions,
$J_1$ and $J_2$, respectively.  There is no frustration along the
$z$ axis (the direction of stacking) and therefore the interaction
can be ferromagnetic or antiferromagnetic along $z$. We will
explain this case in detail since it shows many phenomena
appearing in frustrated systems.

\begin{itemize}
\item{{\bf ${\bf J_2=0}$:}} Consider the case without second
neighbor interaction ($J_2=0$). It is\index{frustration}  not
possible for all three spins at the corners of a triangle to have
the optimal antiparallel orientation which would minimize the
energy of individual pair interactions.  The resulting compromise
in the case of vector spins is the so--called $120^\circ$
structure, as shown in Fig. ~\ref{fig_GS_STA_I} (see chapter 1).
In frustrated systems, local minimization of the energy is not
compatible with the global energy minimum (or minima).  A formal
definition proposed by Toulouse\cite{toul} (and also
Villain\cite{vill1}) in the study of spin glasses states that a
geometry is frustrated if the sign of the product of exchange
interactions $J_i$ around a plaquette $C$
\begin{equation}
\Phi_C = {\rm sign} \bigl[\, \prod_{i \epsilon C} J_i \,\bigr]
\end{equation}
is negative (where $J_i <0 $ implies antiferromagnetic
interactions).  An antiferromagnetic triangular plaquette is thus
frustrated as it involves a product of three $J_i < 0$.  The
triangular lattice is {\it fully frustrated} since all plaquettes
satisfy this rule. The principal effect of the frustration here is
that it gives rise to a non--collinear magnetic order. This
spin-order GS (I) is planar and stable as long as next nearest
neighbor interaction is small, that is $0\leq J_2/J_1\leq 0.125$.

\begin{figure}[th]
\centerline{\psfig{file=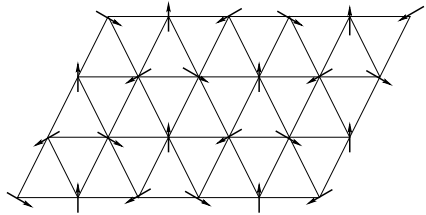,width=2.2in,angle=0}}
\vspace*{8pt}
\caption{
Ground State I of the antiferromagnetic triangular lattice (first neighbors only).
The GS has a $120^\circ$ spin structure.
\label{fig_GS_STA_I}
}
\end{figure}

The BS between the disordered phase at high temperatures and the phase
at low temperatures is $O(N)/O(N-2)$ which is equivalent to
$Z_2\otimes SO(2)$ for $N=2$ and $SO(N)/SO(N-2)$ for $N>2$.
We notice that the exact value of the angle between the spin directions
plays no role for the phase transition.

This system has been extensively studied in three dimensions for
$XY$ spins\cite{Kawa92,PlumerXY,Loison
96,Itakura2001,Plumer_STA_1D} for Heisenberg
spin,\cite{Kawa92,Itakura2001,PlumerHei,LoisonHei,Bhattacharya}
and for the unphysical systems $N=6$ in Ref. [13] and $N=8$ in
Ref. [8]. The two dimensional systems have also been the subject
of several studies for $XY$
spins\cite{JLee91,Miyashita84,LeeLee98,Caprioti97,Loison_TA_XY}
and also for Heisenberg spins.\cite{Wintel_O3_2d}

\item{{\bf ${\bf 0.125 \leq J_2/J_1 \leq 1}$:}} Consider the
presence of an antiferromagnetic second neighbor interactions
($J_2$) in the $xy$ plane. The GS can be determined by minimizing
the energy after a Fourier transform.\cite{Jolicoeur90} For $0
\leq J_2/J_1 \leq 0.125$, the GS is still the $120^\circ$
structure. For $0.125 \leq J_2/J_1 \leq 1$ the GS is degenerate
with  $\theta$ taking any value between 0 and $2\pi$ (see
Fig.~\ref{fig_GS_STA_II}). However the degeneracy will be lifted
by thermal fluctuations (spin waves) and only a collinear GS (II)
will be chosen.\cite{Henley} This phenomenon, called "order by
disorder" following Villain,\cite{vill3}  is general in frustrated
systems and we will see other examples (fcc, hcp) later. There are
three ways to choose the parallel spins and the BS of the lattice
symmetry $C_{3v}$ is a three-state Potts symmetry $Z_3$. The total
BS is $Z_3 \otimes O(N)/O(N-1)$. For $XY$ spins ($N=2$) it is
equivalent to $Z_3\otimes SO(2)$, and to $Z_3 \otimes
SO(N)/SO(N-1)$ for $N>2$.

Numerical studies have been done for the three-dimensional case by
Loison, Diep and Boubcheur.\cite{Loison 96,LoisonHei}

\begin{figure}[th]
\centerline{\psfig{file=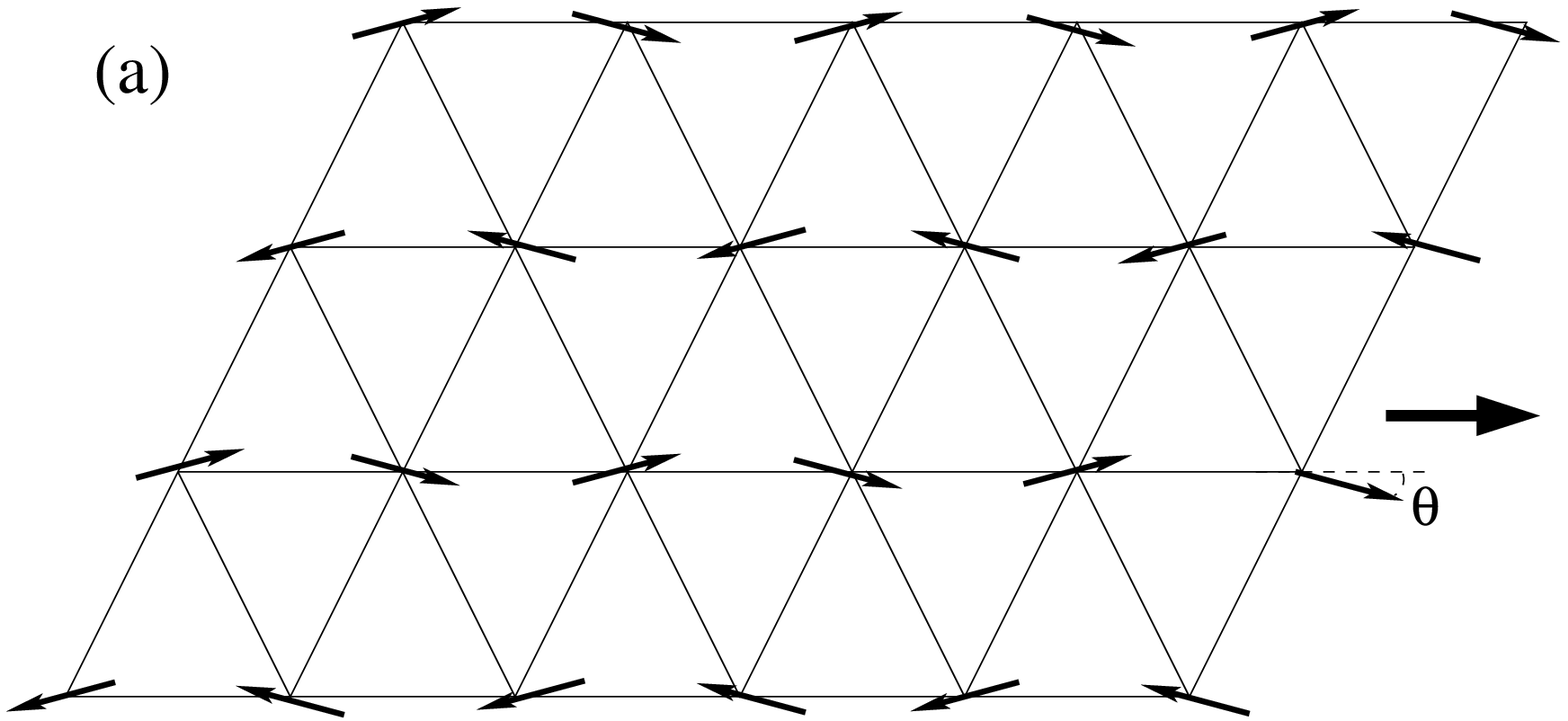,width=2.2in,angle=0}
            \psfig{file=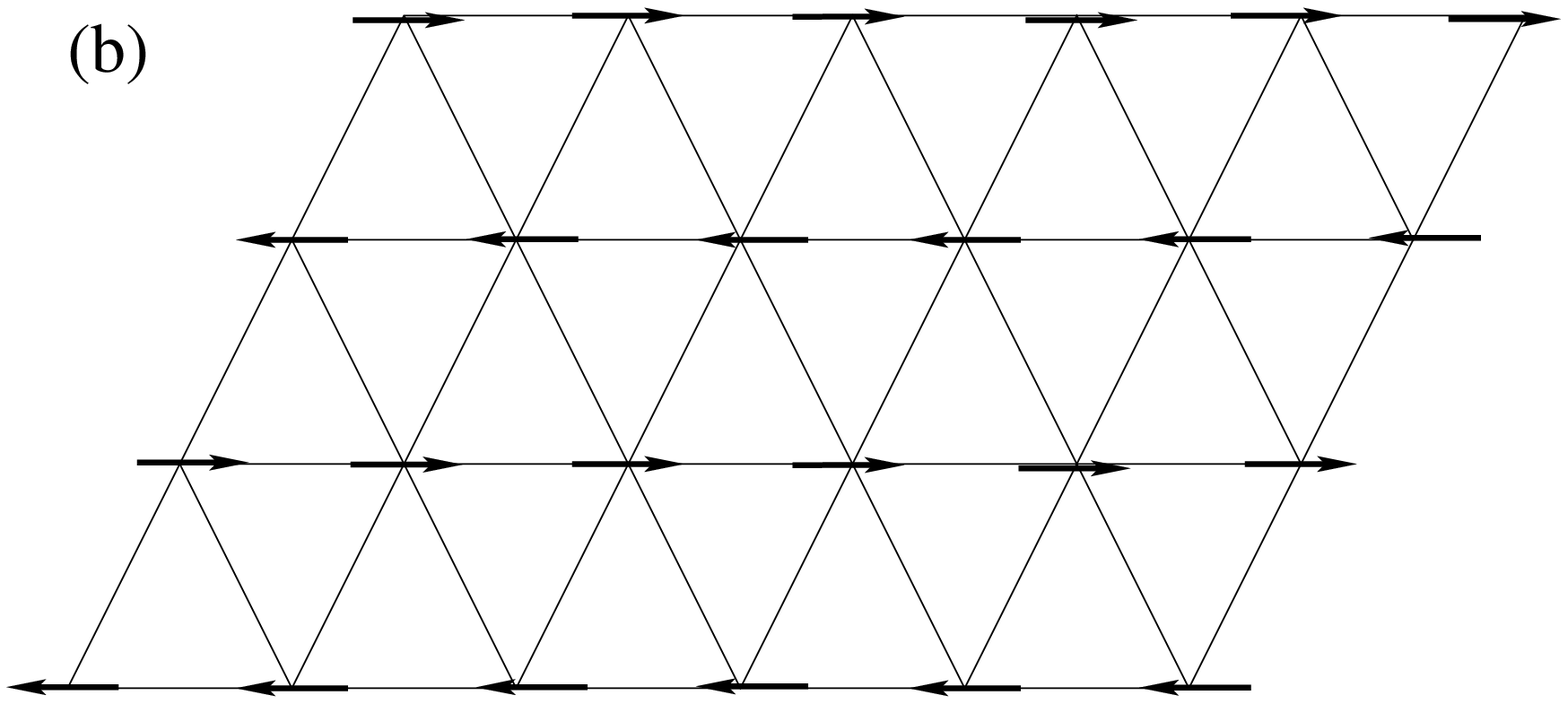,width=2.2in,angle=0}}
\vspace*{8pt} \caption{ Ground State (II) of the antiferromagnetic
triangular lattice with first ($J_1$) and second ($J_2$)-neighbor
antiferromagnetic interactions. $0.125 \leq J_2/J_1 \leq 1$. (a)
$\theta$ can take any value between 0 and $2\pi$. (b) at $T >0$
only the ``most collinear'' GS is chosen. \label{fig_GS_STA_II} }
\end{figure}

\item{{\bf $J_2/J_1>1$}:} Now consider that $J_2/J_1>1$. The
ground state is also degenerate, but this degeneracy is lifted by
thermal fluctuations. The GS (III) is no more collinear, but still
planar (see Fig.~\ref{fig_GS_STA_III}) and there is still three
ways to choose the parallel spins. $\alpha$ is determined by
$\cos(\alpha)=-0.5 \cdot (1+J_1/J_2)$ and can be incommensurable,
i.e. does not correspond to a rational value. Furthermore $\alpha$
varies slowly as a function of the temperature. The BS is $Z_3
\otimes O(N)/O(N-2)$ which is equivalent to $Z_3 \otimes Z_2
\otimes SO(2)$ for $N=2$ ($XY$ spins), to $Z_3 \otimes SO(3)$ for
$N=3$ (Heisenberg spins), and to $Z_3 \otimes SO(N)/SO(N-2)$ if
$N\geq 4$. Numerically the incommensurable angle is problematic.
Indeed we have to impose periodic boundary conditions and since
$\alpha$ varies with the temperature, the size chosen compatible
with $\alpha(T=0)$ is no more compatible with
$\alpha(T>0)<\alpha(T=0)$ for higher temperatures. Therefore the
use of Finite Size Scaling technique (FSS, see \ref{appendixMC})
to calculate the critical exponents will be
problematic.\cite{Saslow92} In addition  the GS III may not be
stable under the new boundary constraint. Then the Potts symmetry
could not be broken and the BS could be just $O(N)/O(N-2)$, and
does not belong to the same universality class as previously. We
will have an akin problem for helimagnets (see later) and for
triangular lattices with two distinct nearest-neighbor
interactions.\cite{Saslow92,STA_assymetric}

This model has been studied in three dimensions by Loison, Diep
and Boubcheur.\cite{Loison 96,LoisonHei}

\begin{figure}[th]
\centerline{\psfig{file=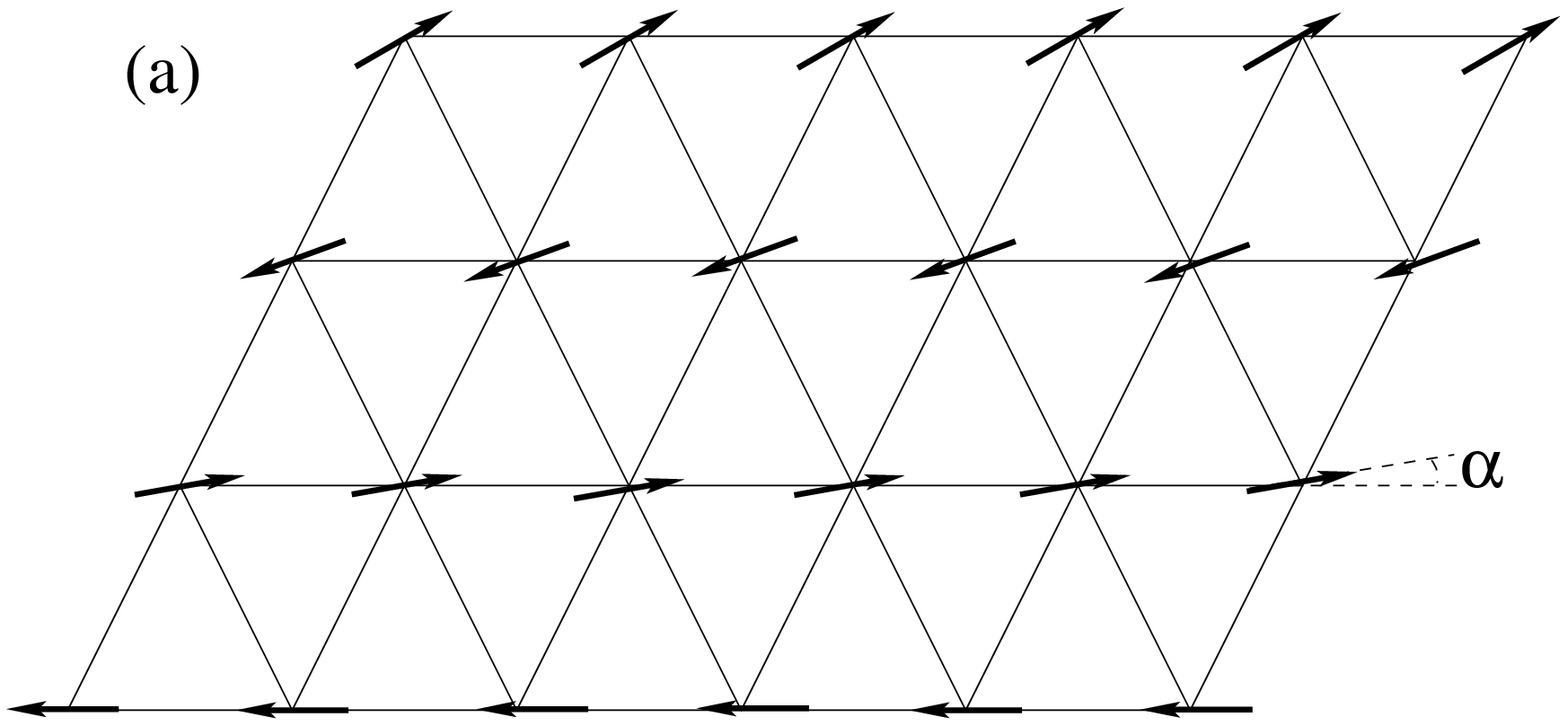,width=2.2in,angle=0}
            \psfig{file=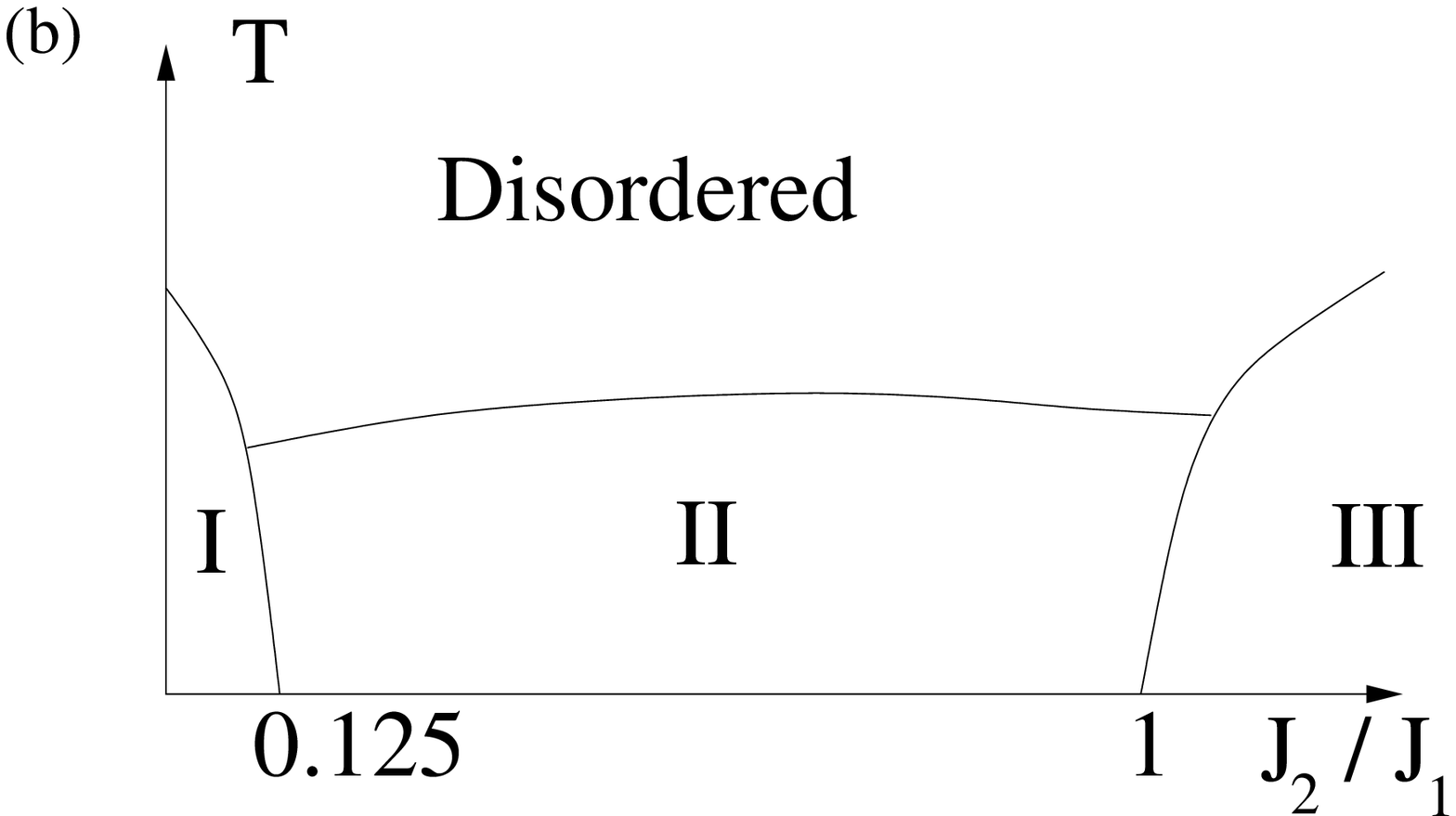,width=2.0in,angle=0}}
\vspace*{8pt} \caption{ (a) Ground State III of the
antiferromagnetic triangular lattice with first ($J_1$) and second
($J_2$) -neighbor antiferromagnetic interactions. $ 1 < J_2/J_1 $.
(b) Phase diagram in the ($J_2/J_1,T$) space.
\label{fig_GS_STA_III} }
\end{figure}

\item{{\bf Other BS:}} We have seen the three BS which appear
between the disordered phase and the GS. In addition two other
transitions appear for a small range of $J_2/J_1$ near 0.125 and
1: between the GS I and II for $0.120 < J_2/J_1 \leq 0.125$, and
between the GS II and III for $1 \leq J_2/J_1 <1.05$.\cite{Loison
96,LoisonHei} This is a general scheme: due to thermal fluctuation
the most collinear state is favored when the temperature
increases. Looking at the symmetries we can get the order and even
the universality class of these transitions.

The GS III and II are compatible in the sense that at the transition $\alpha$ will
go smoothly to zero and the transition could be of second order. The BS between
III and II is $O(N-1)/O(N-2)$  and consequently the transition should be $N-1$ ferromagnetic
type.

The GS I and II are incompatible because there is no way to go smoothly from one to the other
and the transition should be of first order.

\end{itemize}

\subsubsection{bct Helimagnets\index{helimagnets} \index{bct}}
\begin{figure}[th]
\centerline{\psfig{file=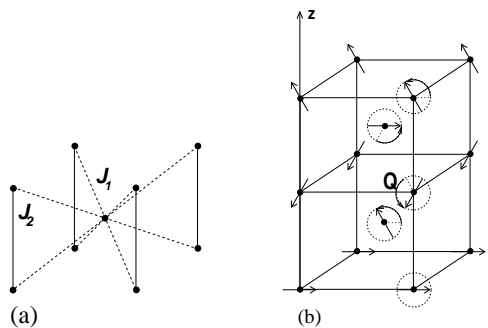,width=2.5in,angle=0}}
\vspace*{8pt}
\caption{
a. Interactions for the bct.
b. Ground State for the bct for $J_2/J_1=0.5$.
\label{fig_GS_Heli}
}
\end{figure}

We consider the body--centered--tetragonal (bct) helimagnets. The
competition between the $J_1$ and $J_2$ interactions gives rise to
a helical ordering along the $z$ axis (see
Fig.~\ref{fig_GS_Heli}). This GS is characterized by the turn
angle $\alpha$ between spins belonging to two adjacent planes
perpendicular to the $z$ axis. $\alpha$ is given by the formula
$\cos(\alpha)=-J_2/J_1$ and $\alpha$ decreases slowly as function
of the temperature. There is no breakdown of the symmetry of the
lattice,  the GS is non collinear but planar and the BS is
$O(N)/O(N-2)$, i.e. identical as the STA with first nearest
neighbor interaction only.

Numerically we have a similar problem  as for the STA with large
NNN interaction: $\alpha(T)$ varies as a function of the
temperature, but with periodic boundary conditions a constraint is
present.

Nevertheless effect of this constraint should not be too strong.
Indeed a small constraint will not break a symmetry of the lattice
and for the $XY$, it will not change the BS of the spin rotation
because all the symmetries of the rotation group are already
broken. For Heisenberg spins the GS could become non coplanar, but
this is unlikely. Therefore we should find the same universality
as that of the STA. The only difference will be a new correction
to the scaling laws for a second order phase transition. But for a
first order one the boundary conditions would not matter.

This model has been studied by Diep  and
Loison\cite{Diep_bct,Loison_bct}. A quantum version has also been
considered by Quartu and Diep \cite{Quartu-Diep_bct}.

\subsubsection{Stacked $J_1$--$J_2$ square\index{J1-J2 square lattice model}
lattices\index{cubic lattice}} 
\begin{figure}[th]
\centerline{\psfig{file=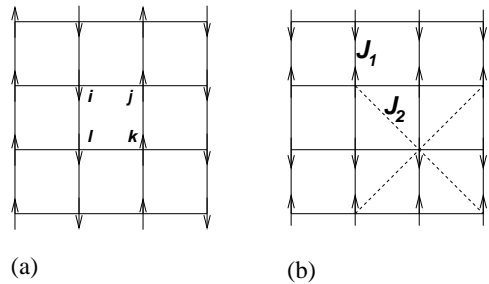,width=2.5in,angle=0}}
\vspace*{8pt} \caption{ Ground State of the stacked $J_1$--$J_2$
lattice with first ($J_1$) and second ($J_2$) -neighbor
antiferromagnetic interactions. $ J_2/J_1 >0.5 $. (a) and (b) show
the two possible configurations at non zero temperature.
\label{fig_GS_J1_J2} }
\end{figure}

The $J_1$--$J_2$ simple cubic
lattice is made by stacking along the $z$ axis the square lattices
with antiferromagnetic (or ferromagnetic) first nearest neighbor
interaction ($J_1$) and antiferromagnetic second nearest neighbor
interaction ($J_2$) in the $xy$ plane. For $ J_2/J_1 >0.5$ the GS
is degenerate but by ``order by disorder'' only collinear
configurations appear (see Fig.~\ref{fig_GS_J1_J2}). There are two
ways to place the parallel spins (following the $x$ or $y$ axis)
and a $Z_2$ Ising symmetry of the lattice symmetry is broken. The
BS will be $Z_2*O(N)/O(N-1)$ or equivalently $Z_2 \otimes SO(2)$
for $N=2$ and $Z_2 \otimes SO(N)/SO(N-1)$ for $N>2$. For $XY$
spins, there exists an identical BS as for the STA with first
nearest neighbor interaction only. Therefore the two systems for
$N=2$ should belong to the same universality class.

No numerical studies have been done on this model for the
three-dimensional case. The two-dimensional case has been studied
by Loison and Simon.\cite{LoisonSimon}

\subsubsection{The simple cubic\index{cubic lattice}  $J_1$--$J_2$ lattice}
The simple cubic $J_1$--$J_2$ lattice is similar to the stacked
$J_1$--$J_2$ square lattices shown above, but the second neighbor
interactions are also present in the $xz$ and $yz$ planes.
Similarly, by ``order by disorder'' the spin configuration is
collinear, but now there are three ways to choose the direction of
parallel spins (the $z$ axis in addition to the $x$ and $y$ axes).
Therefore the BS is $Z_3 \otimes O(N)/O(N-1) \equiv Z_3 \otimes
SO(N)/SO(N-1)$. It is equivalent to the BS of the STA with
intermediate NNN interaction.

This model has been studied for Heisenberg spin by Alonso et
al,\cite{Alonso96} and Pinettes and Diep\cite{Diep_cubic_J1_J2}.

\subsubsection{$J_1$--$J_2$--$J_3$ lattice}
The addition of a third-neighbor antiferromagnetic interaction to
the previous model leads to new GS and therefore to a new BS. For
some value of the interactions a non planar GS state
appears.\cite{Moreo90} In addition a $Z_2$ symmetry is broken from
the lattice group and the BS is hence $Z_2 \otimes O(N)/O(N-2)$
between the ordered phase and the disordered phase. For $XY$ spins
the BS is equivalently $Z_2\otimes Z_2\otimes SO(2)$, and for
Heisenberg spins $Z_2\otimes SO(3)$. There exist also various
transitions between various phases, but following the discussion
at the end of section~\ref{STA_III} concerning STA with NNN
interaction, it is not difficult to find that the transition will
be of first order or of the ferromagnetic type $SO(N-1)/SO(N-2)$.

No numerical studies have been done so far on this model.

\subsubsection{Villain\index{Villain lattice}  lattice and fully
\index{fully frustrated simple cubic lattice} frustrated simple
cubic lattice} 
\begin{figure}[th]
\centerline{\psfig{file=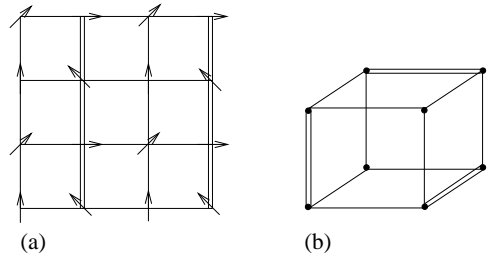,width=2.5in,angle=0}}
\vspace*{8pt} \caption{ a. Ground state of the stacked Villain
lattice. The double lines are antiferromagnetic interactions. b.
The fully frustrated cubic lattice. \label{fig_GS_Villain} }
\end{figure}

The fully frustrated square lattice, called Villain
lattice, is shown in Fig.~\ref{fig_GS_Villain}a. This
two-dimensional lattice has been extensively
studied\cite{vill1,JLee91,Teitel83,Berge86,Nicolaides91,Jose96,Thijssen90,Granato93,LeeLee94,Luo98,Olsson95}.

The fully frustrated simple cubic lattice is made from the Villain
lattice in three dimensions (see Fig.~\ref{fig_GS_Villain}b). The
lattice  has three ferromagnetic interactions and one
antiferromagnetic one for each plaquette. In this case the GS has
an infinite degeneracy for Heisenberg spins and is twelve-fold
degenerate for $XY$ spins. Diep et
al\cite{DiepGhazali,Lallemand-Diep} and Alonso et
al\cite{Alonso96} have studied this model.

\subsubsection{Face-centered cubic lattice (fcc\index{fcc})}
\begin{figure}[th]
\centerline{\psfig{file=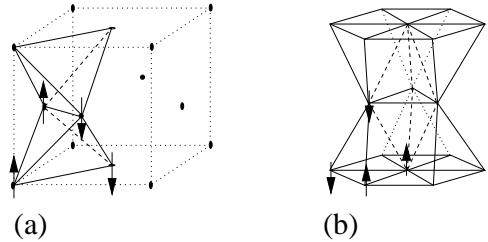,width=2.5in,angle=0}}
\vspace*{8pt}
\caption{
a. Ground State of the fcc lattice.
b. Ground State of the hcp lattice.
\label{fig_GS_fcc}
}
\end{figure}

The face-centered cubic lattice (fcc) has a degenerate GS. But
like the STA with NNN interaction, the ``most collinear phase'' is
selected by thermal fluctuations. If we consider the smaller unit
tetrahedral cell (see Fig.~\ref{fig_GS_fcc}), the GS consists of
two parallel and two antiparallel spin pairs. There are three ways
to construct. This is equivalent to a Potts $Z_3$ symmetry.
Therefore, since the GS is collinear, the BS will be $Z_3 \otimes
O(N)/O(N-1)$ or, equivalently, $Z_3 \otimes SO(N)/SO(N-1)$. This
BS is identical to the STA with intermediate NNN interaction.

This model has been studied by Diep and Kawamura\cite{Diep_fcc}
and Alonso et al.\cite{Alonso96}

\subsubsection{Hexagonal--close--packed lattice (hcp\index{hcp})}
The hexagonal--close--packed lattice (hcp) has many common
features with the fcc. It is constructed by stacking tetrahedra
(see Fig.~\ref{fig_GS_fcc}) and an ``order by disorder'' mechanism
lifts one part of the degeneracy of the GS resulting in a
collinear GS composed with two parallel and two antiparallel
spins. Equivalently to the fcc case,  the BS will be $Z_3 \otimes
O(N)/O(N-1)$ or equivalently $Z_3 \otimes SO(N)/SO(N-1)$. This BS
is identical to the STA with intermediate NNN interaction.

This model has been studied by Diep\cite{Diep_hcp}

\subsubsection{Pyrochlores\index{pyrochlores}}
The pyrochlore lattice can be made by stacking Kagom\'e lattices
along (111) direction and is composed of an arrangement of
corner-sharing tetrahedra (see the chapters by Bramwell et al, and
by Gaulin and Gardner, this book). In presence of the first and
third nearest neighbor interactions Reimers\cite{Reimers2} has
proved that the GS is collinear. Similar to the fcc and hcp cases,
there are three ways to place the spins on each tetrahedron. The
BS should therefore be $Z_3 \otimes O(N)/O(N-1)$ or equivalently
$Z_3 \otimes SO(N)/SO(N-1)$. This BS is identical to the STA with
intermediate NNN interaction.

Reimers et al.\cite{Reimers2} have studied this model.

\subsubsection{Other lattices}

We could stack a Zig--Zag\index{zig-zag model}
model\cite{Boubcheur98} and various phases and phase transitions
should appear. With an equivalent analysis described at the end of
the section~\ref{STA_III} concerning STA with NNN interaction, it
is not difficult to find the nature of the transitions. Still,
other lattices can be considered, however as shown later, they
will have equivalent BS.

\subsubsection{STAR\index{STA}\index{STAR}  lattices}\index{stacked triangular antiferromagnets}

In addition one can construct spin systems having an identical
breakdown of symmetry even if they do not correspond to a real
system.

The first example is derived from the STA. Following the chapter
of Delamotte et al. in this book \cite{chapter_delamotte} certain
modes are irrelevant near the critical point. In particular we can
construct cells of three spins which are always in the ground
state 120$^{\circ}$ configuration. One gets a system of cells in
interaction but with a rigidity imposed inside the cells (see
Fig.~\ref{fig_GS_STAR}). We call this system STAR (Stacked
Triangular Antiferromagnetic with Rigidity). The BS is then
$O(N)/O(N-2)$, identical to STA with NN interaction.

This model has been studied in three dimensions by Dobry and
Diep\cite{Dobry} and by Loison and
Schotte.\cite{LoisonSchotteXY,LoisonSchotteHei,LoisonSchotteO4}

\subsubsection{Dihedral\index{dihedral lattice}  lattices $V_{N,2}$}
\begin{figure}[th]
\centerline{\psfig{file=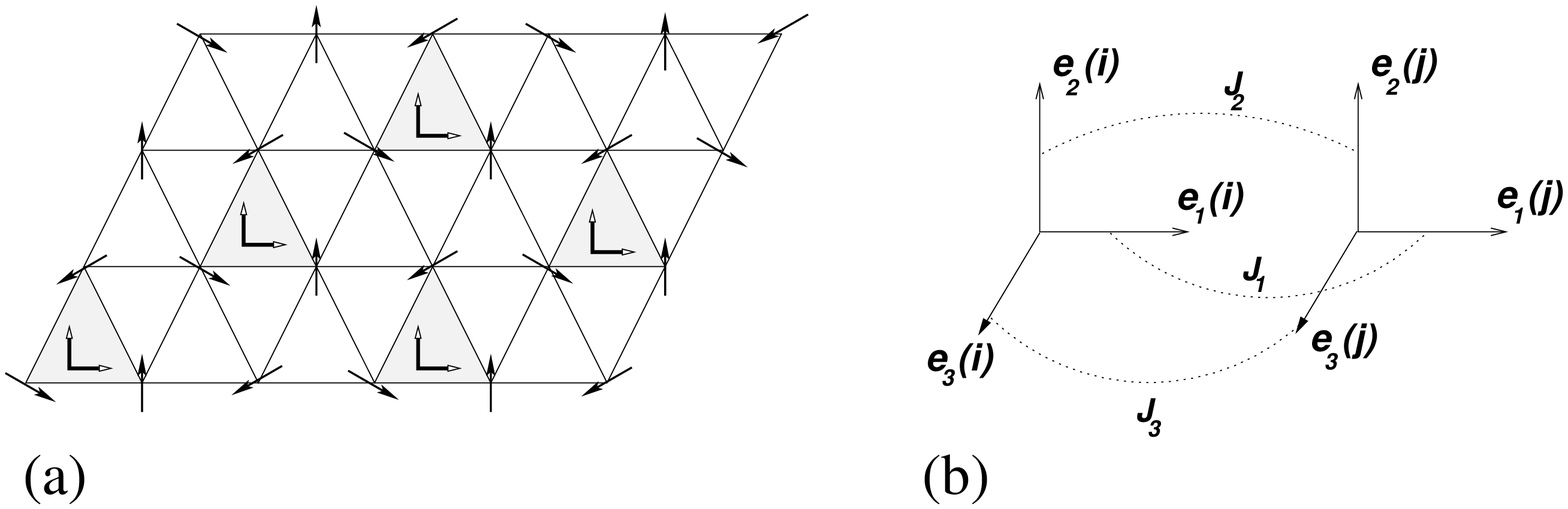,width=4.0in,angle=0}}
\vspace*{8pt}
\caption{
a. STAR model and dihedral model.
b. Trihedral model and $V_{3,P}$ model. $P=1$ for $J_2=J_3=0$, $P=2$ for $J_3=0$ and
$J_1=J_2=1$, and $P=3$ for $J_1=J_2=J_3=1$.
\label{fig_GS_STAR}
}
\end{figure}

The second example is derived from the STAR. It is composed of two
vector spins ${\bf e}_1$ and ${\bf e}_2$ constrained to be
orthogonal to each other at each lattice site. The interactions
are set to be ferromagnetic and the spin ${\bf e}_1$ (${\bf e}_2$)
interacts only with the other spins ${\bf e}_1$ (${\bf e}_2$) at
other sites (see the Fig.~\ref{fig_GS_STAR}). This model is
referred to as the dihedral model $V_{N,2}$. We note that for $XY$
spins $N=2$ the model can be right--handed or left--handed.

The Hamiltonian is defined by
\begin{equation}
\label{H_dihedral}
H = J \sum_{<ij>}  \sum_{k=1}^P
\Big{[} \ {\bf e}_{k}(i)\cdot{\bf e}_{k}(j) \ \Big{]}
\end{equation}
where $P=2$.

At high temperatures the symmetry is $O(N)$ for the first vector
and $O(N-1)$ for the second one, since the two vectors must be
orthogonal. At low temperatures the symmetry is $O(N-1)$ for the
first vector and $O(N-2)$ for the second one for the same reason.
Therefore the BS is $O(N)/O(N-2)$, identical to the STA with NN
interaction.

This model has been studied in three dimensions by Kunz and
Zumbach\cite{Kunz} and by Loison and
Schotte\cite{LoisonSchotteXY,LoisonSchotteHei}, and in two
dimensions for $XY$ spins by Nightingale, Granato, Lee, and
Kosterlitz.\cite{Nightingale95,Granato91}

\subsubsection{Right--handed trihedral\index{trihedral lattice}  lattices $V_{3,3}$}
For Heisenberg spins one can construct another model from the
dihedral $V_{3,2}$ with an identical BS. By adding a third vector
${\bf e}_3={\bf e}_1\times {\bf e}_2$ to the dihedral model no
degree of freedom is added. Therefore the BS is unchanged and
identical to the dihedral $V_{3,2}$ and to the STA with NN
interaction, $O(3)/O(1)\equiv SO(3)$.

This model has been studied in three dimensions by
Loison and Diep.\cite{Loison_triedre}

\subsubsection{P--hedral lattices $V_{N,P}$}

We can generalize the dihedral $V_{N,2}$ model to $V_{N,P}$ with
$P$ orthogonal vectors at each site. For $P=N$ the $N$ vectors can
be right- or left-handed. The BS of symmetry is then $O(N)/O(N-P)$
with $P\leq N$.

For $P=1$ we have the ferromagnetic case, i.e.  a collinear GS.
For $P=2$ we have a coplanar GS but no more collinear like in the
STA with NN interaction. For $P=3$ the GS is no more coplanar but
restricted to a space in three dimensions, and so on. The case
$N=P=3$ could correspond to some experimental systems and spin
glasses should also have this kind of breakdown of symmetry but in
presence of disorder. For $P=N$ the BS is $O(N)/O(0) \equiv Z_2
\otimes SO(N)$.

This model has been studied by Loison.\cite{Loison_V_NP}

\subsubsection{Ising and Potts-$V_{N,1}$ model}

We define the following Hamiltonian
\begin{eqnarray}
\label{H_ON_Potts}
H = -J_1 \sum_{(ij)} {\bf S}_{i}.{\bf S}_{j}.\delta_{q_iq_j}
\end{eqnarray}
where ${\bf S}_{i}$ are  $N$-component classical vectors of unit
length, $\delta_{q_iq_j}$  are the $q$-state Potts spin ($q=2$ for
Ising spin) with $\delta_{q_iq_j}=0$ when $q_i\ne q_j$,  and the
interaction constant $J_1$ is taken positive (ferromagnetic
interaction). The sum runs over all nearest neighbors. For $N=2$
this model is exactly equivalent to the dihedral $V_{2,2}$ model
introduced previously.

Obviously the  BS is  $Z_q \otimes O(N)/O(N-1)$ or equivalently
$Z_q \otimes SO(N)/SO(N-1)$ which is identical to the stacked
$J_1$--$J_2$ model if $q=2$, and to the STA with NNN interaction,
fcc, hcp if $q=3$.

This model has been studied in three dimensions by Loison.\cite{Loison_preparation}
and in two dimensions for $XY$ spins by Nightingale et al.\cite{Nightingale95,Granato91}

\subsubsection{Ising and Potts--$V_{N,2}$ model}
We can also define a dihedral\index{dihedral lattice}  model with
the Hamiltonian (\ref{H_dihedral}) coupled with a Potts model in
the same way as in (\ref{H_ON_Potts}). In this case the BS is $Z_q
\otimes O(N)/O(N-2)$ which is identical to the BS for the stacked
$J_1$--$J_2$--$J_3$ model if $q=2$ and to the STA with a large NNN
for $q=3$.

This model has been studied by Loison.\cite{Loison_preparation}

\subsubsection{Landau\index{Landau--Ginzburg}--Ginzburg model}

The Landau--Ginzburg Hamiltonian is constructed from the dihedral
$V_{N,2}$ model (see chapter of Delamotte et
al.\cite{chapter_delamotte}). We release the constraints on the
orthogonality of the spins and on the norm unity, replacing them
by a potential we arrive at the following Hamiltonian
\begin{eqnarray}
\label{hamiltonian_uns_dihe}
H={K} \sum_{<ij>}
(\vec{\phi}(i)-\vec{\phi}(j))^2
+r\sum_i
\vec{\phi}(i)^2
+u \sum_i
\left(\vec{\phi}(i)^2 \right)^2  \notag
\\
+v \sum_i
\left[
(\vec{\phi}_a(i) \cdot \vec{\phi}_b(i))^2 -
\vec{\phi}_a(i)^2 \vec{\phi}_b(i)^2
\right]
\end{eqnarray}
where $\vec{\phi}(i)=\left(\vec{\phi}_a(i), \vec{\phi}_b(i)
\right)$ is an $N+N$--component vector defined on the lattice site
i, and the summation $\sum_{<ij>}$ runs over all nearest neighbor
pairs of the lattice. The first term represents the interactions
between the sites and the last term the constraint that the spins
$\phi_a(i)$ and $\phi_b(i)$ are orthogonal. If $v=0$, the
Hamiltonian is reduced to a standard $O(2N)$ ferromagnetic
problem. This is our third model that we call the chiral $\phi^4$
model. For $v>0$ the BS is $O(N)/O(N-2)$, identical to STA with NN
interaction.

This model has been studied numerically by Itakura.\cite{Itakura2001}

\subsubsection{Cubic\index{cubic term} term in Hamiltonian}
One can introduce a model where the frustration is not geometrical
but included on each link resulting in a non collinear GS.
Consider the Hamiltonian for two spins:
\begin{equation}
\label{eq1}
H=J_1 {\bf S}_{i}.{\bf S}_{j}
+ J_3 ({\bf S}_{i}.{\bf S}_{j})^3
\end{equation}
where ${\bf S}_{i}$ is a $N$ component classical vector of unit length,
$J_1$ is a ferromagnetic coupling constants $J_1<0$
 and $J_3$ an antiferromagnetic one $J_3>0$.
This cubic term has the same symmetries as the linear term and for
a ferromagnetic system ($J_3<0$) the universality class will be
identical.\cite{Loison_SS3} For $1/3 \le -J_3/J_1 \le 4/3$ the
minimum of $H$ occurs when the two spins are canted with an angle
$\cos(\alpha)=\frac{1}{\sqrt{3 J_3/J_1}}$. On a cubic lattice the
GS is planar, on a triangular lattice it is in three dimensions
(if $N\ge3$), on a fcc lattice it is in four dimensions (if
$N\ge4$), $\cdots$ This phase has, at finite temperature, a
transition to a ferromagnetic collinear phase. Therefore the BS
will be $O(N-1)/O(N-P)$ with $P=2$ for the cubic lattice (planar
GS), $P=3$ for the stacked triangular, $P=4$ for the fcc lattice
$\cdots$ The case $P=2$ has been tested for Heisenberg spins and
it gives indeed an $O(2)/O(1) \equiv SO(2)$ transition, i.e. a
ferromagnetic $XY$ transition.\cite{Loison_SS3} The case $P=3$ has
an identical BS as the STA but it has not been tested.

This model has been studied numerically by Loison.\cite{Loison_SS3}

\subsubsection{Summary}

The number of models that can be constructed by adding
interactions is unlimited. However, the number of BS can only be
finite. As has been shown in the previous sections, the symmetry
breaking is limited to $Z_q\otimes O(N)/O(N-P)$.  For the
physically relevant cases it is restricted to $q=1$ (Identity),
$q=2$ (Ising), or $q=3$ (Potts), $N=2$ ($XY$ spins), $N=3$
(Heisenberg spins), and $P=1,\,2$ or 3. In total for Heisenberg
spins there will exist 8 cases (two are identical: $(q=1,P=3)
\equiv (q=2,P=2)$) and 5 for $XY$ spins.

The tables below summarize the probable BS for physical frustrated
systems with $XY$ or Heisenberg spins.

\begin{table}
\tbl{Most probable BS for frustrated $XY$\index{XY spin} spin
systems with corresponding lattices or models \index{universality
class}\index{ferromagnets} \index{critical
exponents}\index{STA}\index{triangular antiferromagnetic lattice}
\index{helimagnets}\index{bct} \index{cubic lattice}\index{Villain
lattice} \index{fully frustrated lattice}\index{fcc}\index{hcp}
\index{pyrochlores}\index{zig-zag model}
\index{STAR}\index{dihedral lattice} \index{cubic
term}\index{breakdown of symmetry} } {\tabcolsep10pt
\begin{tabular}{@{}c|@{\hspace*{0.7cm}}c}
\hline\\[-3pt]
{\bf BS}& {\bf lattice-model} \\[2pt]
\hline\\[-8pt]
$SO(2)$ & ferromagnetic, small frustration   \\[2pt]
\hline\\[-8pt]
$Z_2 \otimes SO(2)$ & STA, STAR, $V_{2,2}\equiv$ Ising-$V_{2,1}$,
bct,\\&Stacked $J_1$--$J_2$, Stacked Villain,\\&Fully Frustrated cubic,
Stacked Zig-Zag,\\& chiral $\phi^4$ model, $(S_i.S_j)^3$ term  \\[2pt]
\hline\\[-8pt]
$Z_3 \otimes SO(2)$ &STA+NNN, cubic $J_1$--$J_2$, fcc, hcp, pyrochlore,
Potts-$V_{2,1}$\\[2pt]
\hline\\[-8pt]
$Z_2 \otimes Z_2 \otimes SO(2)$ &stacked $J_1$--$J_2$--$J_3$, Ising-$V_{2,2}$\\[2pt]
\hline\\[-8pt]
$Z_3 \otimes Z_2 \otimes SO(2)$ &STA+NNN, Potts-$V_{2,2}$\\[2pt]
\hline
\end{tabular}}
\label{table_BS_O2}
\end{table}

\begin{table}
\tbl{Most probable BS for frustrated Heisenberg\index{Heisenberg
spin} spin systems with corresponding lattices or models
\index{universality class}\index{ferromagnets} \index{critical
exponents}\index{STA}\index{helimagnets}\index{bct} \index{cubic
lattice}\index{Villain lattice} \index{fully frustrated
lattice}\index{fcc}\index{hcp} \index{pyrochlores}\index{zig-zag
model} \index{STAR}\index{dihedral lattice} \index{trihedral
lattice}\index{cubic term}\index{breakdown of symmetry} }
{\tabcolsep10pt
\begin{tabular}{@{}c|@{\hspace*{0.7cm}}c}
\hline\\[-3pt]
{\bf BS}& {\bf lattice--model} \\[2pt]
\hline\\[-8pt]
$SO(3)/SO(2)$ & ferromagnetic, small frustration \\[2pt]
\hline\\[-8pt]
$SO(3)$ & STA, STAR, $V_{3,2}$, right--handed trihedral, bct,\\&Stacked Villain,
Stacked Zig--Zag, \\&Fully Frustrated cubic(?), \\& chiral $\phi^4$ model,
$(S_i.S_j)^3$ term   \\[2pt]
\hline\\[-8pt]
$Z_2 \otimes SO(3)/SO(2)$ & Stacked $J_1$--$J_2$, Ising--$V_{3,1}$   \\[2pt]
\hline\\[-8pt]
$Z_2 \otimes SO(3)$ & Stacked $J_1$--$J_2$--$J_3$, $V_{3,3}\equiv$Ising--$V_{3,2}$,\\&
$(S_i.S_j)^3$ term, Fully Frustrated cubic(?) \\[2pt]
\hline\\[-8pt]
$Z_2 \otimes Z_2 \otimes SO(3)$ & Ising--$V_{3,3}$ \\[2pt]
\hline\\[-8pt]
$Z_3 \otimes SO(3)/SO(2)$ & STA+NNN, cubic $J_1$--$J_2$, fcc, hcp, pyrochlore,
Potts--$V_{3,1}$   \\[2pt]
\hline\\[-8pt]
$Z_3 \otimes SO(3)$ & STA+NNN, Potts--$V_{3,2}$   \\[2pt]
\hline\\[-8pt]
$Z_3 \otimes Z_2 \otimes SO(3)$ & Potts--$V_{3,3}$ \\[2pt]
\hline
\end{tabular}}
\label{table_BS_O3}
\end{table}

\section{Phase transitions between two and four dimensions: $2<d \leq 4$}

In this section we will concentrate on the nature of the various
transitions mentioned in the previous section. Especially the
transition for the $O(N)/O(N-2)$ BS is considered in detail since
it appears in numerous systems and was extensively debated. Then
we will discuss the other BS which are less problematic.

\subsection{$O(N)/O(N-2)$ breakdown of symmetry}

\subsubsection{Fixed\index{fixed point} points}
\begin{figure}[th]
\centerline{\psfig{file=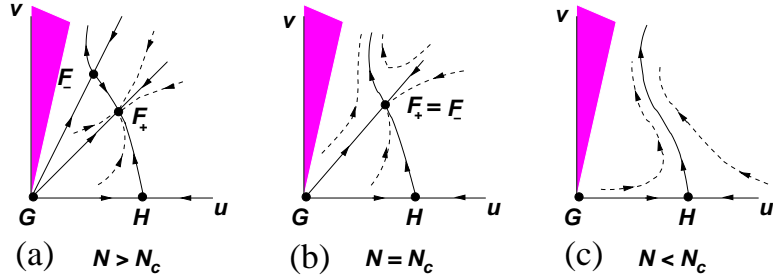,width=4.0in,angle=0}}
\vspace*{8pt}
\caption{
(a), (b), (c) Hamiltonian flow induced by
renormalization\index{renormalization group} group transformations. The
arrows indicate the direction of flow under iterations.
\label{fig_RG_tri}
}
\end{figure}

Since many models (STA, STAR, dihedral, chiral $\phi^4$ model,
$\cdots$, see Tables~\ref{table_BS_O2}-~\ref{table_BS_O3}) have an
identical BS, they should have equivalent critical behavior, i.e.
they belong to the same universality class. However, the situation
is more complicated and even with an identical BS two systems
could show different behavior. To understand this fact we have
plotted in Fig.~\ref{fig_RG_tri} the fixed points in the critical
plan for this model. Since there are two fields
(eq.~\ref{hamiltonian_uns_dihe}) to allow coplanar non collinear
GS we will have four possible fixed points. There are two more
than the ferromagnetic case which has a collinear GS and
consequently only one field.

The fixed points are:
\begin{itemize}
\item[1.] The Gaussian fixed G point at $u^* = v^* = 0$ with
mean-field critical exponents. \item[2.] The $O(2N)$ fixed point H
at $v^* = 0$ and $u^* = u_H\ne 0$ with $O(2N)$ exponents (see
Table \ref{table_ferro}). \item[3.]Two fixed points $F_+$ and
$F_-$ at location ($u_{F_+},v_{F_+})$ and $(u_{F_-},v_{F_-}$)
different from zero. These are the fixed points associated with a
new universality class.
\end{itemize}
The existence and stability of the fixed points depend on the number
of components $N$:
\begin{itemize}
\item[a.]  $N>N_c$: four fixed points are present but three are
unstable ($G$, $H$, $F_-$) and a stable one $F_+$. Therefore the
transition belongs to a new universality class different from the
standard $SO(N)/SO(N-1)$ class. If the initial point for the RG
flow is to the left of the line ($G,\,F_-$), see
Fig.~\ref{fig_RG_tri}a, the flow is unstable and the transition
will be of first order. Therefore two systems with the same
Hamiltonian and the same breakdown of symmetry could have
different critical behaviors.
\item[b.] $N=N_c$: the fixed points $F_-$ and $F_+$ coalesce to a
marginally stable fixed point. One would think that the transition
is ``tricritical'' but the exponents are different and not given
by the tricritical mean-field values contrary to common belief.
The reason is that there are two non zero quartic coupling
constants, see Fig.~\ref{fig_RG_tri}b, in contrast to the
``standard'' tricritical point where the quartic term disappears
and a sextic term takes over.
\item[c.] $N<N_c$: F$_-$ and $F_+$ move into the complex parameter
space, see Fig.~\ref{fig_RG_tri}c. The absence of stable fixed
points is interpreted as a signature of a first-order transition.
\end{itemize}
There are at least three questions: the value of $N_c$,
the location of the initial point in the $RG$ flow, and the nature of the transition.

\subsubsection{MCRG and first-order transition}
The most reliable answers to these questions can be found in the
article of Itakura.\cite{Itakura2001} He studied the
STA\index{stacked triangular antiferromagnets}\index{STA}  model,
dihedral\index{dihedral lattice} model, and the chiral $\phi^4$
model with the Hamiltonian (\ref{hamiltonian_uns_dihe}), using
Monte Carlo Renormalization Group (MCRG). This method is very
powerful to give the flow diagram but not the critical exponents
with a great precision.

He showed that the  STA and dihedral models have an initial point
in the RG flow under the line $GF_-$ in the Fig.~\ref{fig_RG_tri}.
Therefore they should belong to the same universality class
provided the fixed point $F_+$ exists, i.e. $N>N_c$. He found that
$N_c$ is between 3 and 8  in three dimensions which means that the
real physical systems\index{XY spin}  $XY$ ($N=2$) and Heisenberg
spins ($N=3$) have a first-order transition.

In addition he did a standard canonical Monte Carlo (MC)
simulation for $XY$ spins on the STA lattice for very large sizes
of $96^3$ and $126^3$. He found a first-order transition in
agreement with MCRG study, contrary to  smaller size systems which
seem to have a second-order
transition.\cite{Kawa92,PlumerXY,Loison 96} For Heisenberg spins
he could not find a first-order transition but using the results
of the MCRG he concludes that the first-order transition could
only be seen for a size larger than $800^3$ which is not
accessible for actual computer resources. For the dihedral model
with a canonical MC he found a clear first-order transition for
large sizes for Heisenberg spins. For $XY$ spins Loison and
Schotte have already shown that the transition is of first order.
Diep\cite{Diep_bct} in 1989 was the first to find a first-order
transition in helimagnetics bct lattice with $XY$ spins. This
result was considered not conclusive because of problems of
periodic boundary conditions in numerical simulation. However, as
noted above, this is not relevant for $XY$ spins (contrary to
Heisenberg spins) and whence this conclusion is indeed correct for
this BS.

We have now to address the problem why phase transitions appear
continuous for small sizes but show the true first-order nature
only for larger sizes. This phenomenon is not restricted to
frustrated spin systems such as  STA but appears also in the
weakly first-order transition of the two-dimensional Potts model
with $q=5$ components.\cite{Baxter,Peczak_Potts,Loison_potts} More
generally it appears when two fixed points collapse and disappears
following one variable (like $F_-$ and $F_+$ in
Fig.~\ref{fig_RG_tri}). See later for a comparison with the Potts
model.

\subsubsection{Complex fixed\index{fixed point} point or minimum in the flow}
\begin{figure}[th]
\centerline{\psfig{file=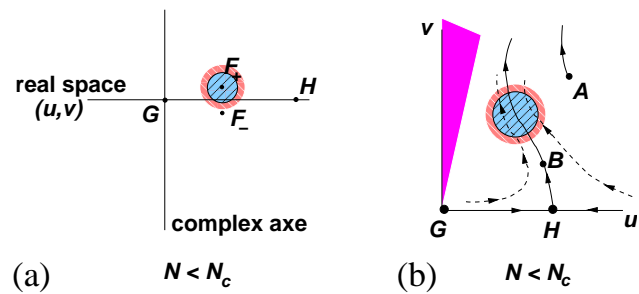,height=1.5in,angle=0}}
\vspace*{8pt} \caption{ a) For $N<N_c$ the fixed points $F_+$ and
$F_-$ become complex. b) Hypothesis on the Hamiltonian flow
induced by renormalization--group transformations for $N<N_c$ (see
Fig.~\ref{fig_RG_tri}c). The arrows give the direction of flow.
The two circles correspond to a low velocity region. The inner one
corresponds to a minimum and hence an ``almost'' second-order
transition. The outer circle corresponds to a slow crossover where
the critical exponents will vary as a function of the system size.
\label{fig_RG_tri2} }
\end{figure}

The change from continuous to discontinuous transition can be
understood using the concept of ``complex fixed point'' or
``minimum in the flow'' first introduced by
Zumbach\cite{Zumbach93} and then developed by Loison and
Schotte.\cite{LoisonSchotteXY,LoisonSchotteHei} In
Fig.~\ref{fig_RG_tri}c for $N<N_c$ the fixed points $F_-$ and
$F_+$ have collapsed and no solution exists in real parameter
space. Instead there exists an imaginary solution plotted in
Fig.~\ref{fig_RG_tri2}a. These solutions should have an influence
on the flow in the real plane as shown in Fig.~\ref{fig_RG_tri2}b.
Zumbach showed that there exists a basin of attraction due to the
complex fixed points where the RG flow is very slow.

If the RG flow of a system goes through this region, the number of
RG iterations $L\rightarrow L/b$ to get outside  will be large.
Therefore the size $L$ of the numerical system considered, should
be made large enough (see \ref{appendixRG}). If the size is ``too
small'', the flow is trapped inside of this low velocity region
and the transition seems to be continuous. We immediately see that
the size $L$ for which the true first-order transition is visible,
that is when the flow is outside of the basin of attraction, will
depend on the starting point. If it is located outside of the
domain of low velocity and if the flow does not approach this
region (point $A$ in Fig.~\ref{fig_RG_tri2}), then the first-order
nature of transition can be seen already for small sizes. In the
other case, if the initial position is such that the flow has to
go through the entire region of low velocity (point $B$ in
Fig.~\ref{fig_RG_tri2}), then the necessary size to reach the true
first-order region will be very large. This explains why STA,
STAR, dihedral and right--handed trihedral\index{trihedral
lattice} (for $N=3$) models having an identical breakdown of
symmetry behave differently. Similarly a single system can show
different behaviors for different sizes.
For the Heisenberg\index{Heisenberg spin}  case, for example, the
right--handed trihedral\index{trihedral lattice}  model shows a
strong first-order behavior even for small sizes. The
dihedral\index{dihedral lattice}  model with the same BS for $N=3$
has a weaker first-order behavior visible only for sizes bigger
than $80^3$ spins while the STA or bct helimagnets show a
second-order transition for similar sizes.

Furthermore the size of this region of low velocity will vary as a
function of the distance between the complex fixed point $F_+$ and
the real plane shown in Fig.~\ref{fig_RG_tri2}a.
It has been shown that $N_c$, where the two fixed points $F_-$ and
$F_+$ collapse and become complex, is bigger than 3 and the
distance of $F_+$ to the real plane will be larger for $N=2$ than
for $N=3$. Therefore the size of the domain of low velocity will
be bigger for $N=3$ than for $N=2$, and the true first-order
behavior will be seen for smaller sizes in the $XY$ case ($N=2$)
than in the Heisenberg case ($N=3$).
Indeed the first-order transition in the dihedral, STAR and STA
can be seen from the size 12, 18 and 96, respectively, for $XY$
spin systems (see Table~\ref{table_O2}). For Heisenberg spins this
has been seen  for the dihedral model with a size $L=80$ (see
Table~\ref{table_O3}) while the estimate size for the STA should
be larger than 800.\cite{Itakura2001} We note that a first-order
transition has been found in quasi--one--dimensional
STA\cite{Plumer_STA_1D}: the initial point in the flow diagram is
outside the domain of low velocity.

For the $XY$ case Delamotte et al.\cite{Tissier} using a non
perturbative RG approach, found that there is no minimum in the
flow but always a low velocity region (outer circle of
Fig.~\ref{fig_RG_tri2}b). Following the initial point in the RG
flow the systems will show  different sets of exponents, i.e.\ the
system is in a crossover region. Numerical simulations tend to
support this interpretation. Indeed for $XY$ spins the STAR and
the STA  for small sizes ($L<40$) display a second-order
transition with different critical exponents (see
Tables~\ref{table_O2}-\ref{table_O2K}) and for large sizes the STA
shows a first-order transition as discussed
above.\cite{Itakura2001}

In conclusion the transition for Heisenberg spins is of first
order but an ``almost second-order transition'' could exist for a
range of ``finite'' sizes. This ``almost second-order transition''
will have a set of exponents different from the ferromagnetic ones
because the breakdown of symmetry is different: $O(N)/O(N-2)$  in
comparison to $O(N)/O(N-1)$ for ferromagnetic systems. This
``new'' universality class has been called chiral
class.\cite{Kawa92} For the $XY$ case there is a crossover region
with a slow velocity and exponents will vary between those of the
chiral class and of weak first-order transitions ($\nu=1/3$ \dots
- see Table~\ref{table_ferro}). The critical exponents of this
chiral class are given in Tables~\ref{table_O2}-\ref{table_O6} for
$N$ varying from 2 to 6 for the STA, STAR, and
dihedral\index{dihedral lattice}  models. With numerical
simulations, exponents $\gamma/\nu$ $\beta/\nu$ and $\nu$ are
usually calculated using finite size scaling while other exponents
are calculated using the scaling relation $d\nu=2-\alpha$ and
$\gamma/\nu=2-\eta$ (see \ref{appendixMC}).

There is another indication given by Loison and
Schotte\cite{LoisonSchotteXY} which suggests that the phase
transition is indeed due of a ``complex fixed point'. Using the
scaling formula $\gamma/\nu=2-\eta$, one gets a negative $\eta$
for $N=2$ and $N=3$ for the STA, STAR or dihedral model (see
Tables~\ref{table_O2}-\ref{table_O6})  which contradicts the
theorem that $\eta$ must be always
positive.\cite{Patashinskii,Zinn89} Negative $\eta$ can also be
calculated using experimental results.\cite{Tissier} Therefore the
fixed point  cannot be real. Using this indication we can conclude
that the continuous transition found in fully frustrated $XY$
lattice\cite{DiepGhazali} is indeed of first order.

\begin{table}
\tbl{ Critical\index{critical exponents} \index{XY spin} exponents
associated to the $SO(2)$ symmetry by Monte Carlo for $XY$ spins
($N=2$) and  a BS $Z_2 \otimes SO(2)$. $^{(a)}$Calculated by
$\gamma/\nu=2-\eta$. The first result{\protect\cite{Kawa92}} comes
from a study at high and low temperatures and uses the FSS. The
second{\protect\cite{PlumerXY}} uses the Binder parameter to find
$T_c$ and uses the FSS, the third{\protect\cite{Loison 96}} uses
the maxima in FSS region. \label{table_O2} } {\tabcolsep10pt
\begin{tabular}{@{}c|@{\hspace*{0.1cm}}c|@{\hspace*{0.1cm}}c|@{\hspace*{0.1cm}}c|@{\hspace*{0.1cm}}c|@{\hspace*{0.1cm}}c|@{\hspace*{0.1cm}}c|@{\hspace*{0.1cm}}c}
\hline\\[-3pt]
system&Ref.&$L_{max}$&$\alpha$&$\beta$&$\gamma$&$\nu$&$\eta$\\[2pt]
\hline\\[-8pt]
STA&\cite{Kawa92}&60&0.34(6)&0.253(10)&1.13(5)&0.54(2)&-0.09(8)$^{(a)}$\\[2pt]
\hline\\[-8pt]
STA\index{STA}&\cite{PlumerXY}&33&0.46(10)&0.24(2)&1.03(4)&0.50(1)&-0.06(4)$^{(a)}$ \\[2pt]
\hline\\[-8pt]
STA&\cite{Loison 96}&36&0.43(10)&&&0.48(2)&\\[2pt]
\hline\\[-8pt]
STA&\cite{Itakura2001}&126&\multicolumn{5}{@{}c@{}} {first order}\\[2pt]
\hline\\[-8pt]
STA&\cite{Plumer_STA_1D}&35&\multicolumn{5}{@{}c@{}} {first order}\\[2pt]
\hline\\[-8pt]
bct\index{bct}&\cite{Diep_bct}&24&\multicolumn{5}{@{}c@{}} {first order}\\[2pt]
\hline\\[-8pt]
STAR\index{STAR}&\cite{LoisonSchotteXY}&36&\multicolumn{5}{@{}c@{}} {first order}\\[2pt]
\hline\\[-8pt]
$V_{2,2}$&\cite{LoisonSchotteXY}&36&\multicolumn{5}{@{}c@{}} {first order}\\[2pt]
\hline
\end{tabular}}
\end{table}

\begin{table}
\tbl{ Critical exponents \index{critical exponents} \index{XY
spin} associated to the $Z_2$ symmetry (chirality $\kappa$) by
Monte Carlo for $XY$ spins ($N=2$) and  a BS $Z_2 \otimes SO(2)$.
$^{(a)}$Calculated by $\gamma/\nu=2-\eta$. The first result
{\protect\cite{Kawa92}} comes from a study at high and low
temperature and uses of FSS. The second {\protect\cite{PlumerXY}}
uses the Binder parameter to find $T_c$ and uses the FSS.
\label{table_O2K} } {\tabcolsep10pt
\begin{tabular}{@{}c|@{\hspace*{0.1cm}}c|@{\hspace*{0.1cm}}c|@{\hspace*{0.1cm}}c|@{\hspace*{0.1cm}}c|@{\hspace*{0.1cm}}c|@{\hspace*{0.1cm}}c|@{\hspace*{0.1cm}}c}
\hline\\[-3pt]
system&Ref.&$L_{max}$&$\alpha$&$\beta_\kappa$&$\gamma_\kappa$&$\nu_\kappa$&$\eta_\kappa$\\[2pt]
\hline\\[-8pt]
STA\index{STA}&\cite{Kawa92}&60&0.34(6)&0.55(4)&0.72(8)&0.60(3)&0.80(19)$^{(a)}$\\[2pt]
\hline\\[-8pt]
STA&\cite{PlumerXY}&33&0.46(10)&0.38(1)&0.90(2)&0.55(1)&0.28(3)$^{(a)}$ \\[2pt]
\hline\\[-8pt]
STA&\cite{Itakura2001}&126&\multicolumn{5}{@{}c@{}} {first order}\\[2pt]
\hline\\[-8pt]
STA&\cite{Plumer_STA_1D}&35&\multicolumn{5}{@{}c@{}} {first order}\\[2pt]
\hline\\[-8pt]
bct\index{bct}&\cite{Diep_bct}&24&\multicolumn{5}{@{}c@{}} {first order}\\[2pt]
\hline\\[-8pt]
STAR\index{STAR}&\cite{LoisonSchotteXY}&36&\multicolumn{5}{@{}c@{}} {first order}\\[2pt]
\hline\\[-8pt]
$V_{2,2}$&\cite{LoisonSchotteXY}&36&\multicolumn{5}{@{}c@{}} {first order}\\[2pt]
\hline
\end{tabular}}
\end{table}

\begin{table}
\tbl{ Critical exponents \index{critical exponents}
\index{Heisenberg spin} by Monte Carlo for Heisenberg spins
($N=3$) and a BS $SO(3)$. Calculated by
$^{(a)}\;\gamma/\nu=2-\eta\,$, $^{(b)}\;d \nu=2-\alpha\,$,
$^{(c)}\;2\,\beta/\nu=d-2+\eta\,$. } {\tabcolsep8pt
\begin{tabular}{@{}c|@{\hspace*{0.1cm}}c|@{\hspace*{0.1cm}}c|@{\hspace*{0.1cm}}c|@{\hspace*{0.1cm}}c|@{\hspace*{0.1cm}}c|@{\hspace*{0.1cm}}c|@{\hspace*{0.1cm}}c}
\hline\\[-3pt]
system&Ref.&$L_{max}$&$\alpha$&$\beta$&$\gamma$&$\nu$&$\eta$\\[2pt]
\hline\\[-8pt]
STA\index{STA}&\cite{Kawa92}&60&0.240(80)&0.300(20)&1.170(70)&0.590(20)&+0.020(180)$^{(a)}$\\[2pt]
\hline\\[-8pt]
STA&\cite{PlumerHei}&36&0.242(24)$^{(b)}$&0.285(11)&1.185(3)&0.586(8)&-0.033(19)$^{(a)}$\\[2pt]
\hline\\[-8pt]
STA&\cite{Bhattacharya}&48&0.245(27)$^{(b)}$&0.289(15)&1.176(26)&0.585(9) &-0.011(14)$^{(a)}$\\[2pt]
\hline\\[-8pt]
STA&\cite{LoisonHei}&36&0.230(30)$^{(b)}$&0.280(15)&&0.590(10)&0.000(40)$^{(c)}$\\[2pt]
\hline\\[-8pt]
bct\index{bct}&\cite{Loison_bct}&42&0.287(30)$^{(b)}$&0.247(10)&1.217(32)&0.571(10)&-0.131(18)$^{(a)}$\\[2pt]
\hline\\[-8pt]
STAR\index{STAR}&\cite{LoisonSchotteHei}&42&0.488(30)$^{(b)}$&0.221(9)&1.074(29)&0.504(10)&-0.131(13)$^{(a)}$ \\[2pt]
\hline\\[-8pt]
$V_{3,2}$ \index{dihedral lattice}  &\cite{LoisonSchotteHei}&40&0.479(24)$^{(b)}$&0.193(4)&1.136(23)&0.507(8)&-0.240(10)$^{(a)}$ \\[2pt]
\hline\\[-8pt]
$V_{3,2}$&\cite{Itakura2001}&80&\multicolumn{5}{@{}c@{}} {first order}\\[2pt]
\hline
\end{tabular}}
\label{table_O3}
\end{table}

\begin{table}
\tbl{ Critical exponents \index{critical exponents} by Monte Carlo
for  spins with four components ($N=4$) and a BS $O(4)/O(2)\equiv
SO(4)/SO(2)$. Calculated by $^{(a)}\;\gamma/\nu=2-\eta\,$,
$^{(b)}\;d \nu=2-\alpha\,$. } {\tabcolsep8pt
\begin{tabular}{@{}c|@{\hspace*{0.1cm}}c|@{\hspace*{0.1cm}}c|@{\hspace*{0.1cm}}c|@{\hspace*{0.1cm}}c|@{\hspace*{0.1cm}}c|@{\hspace*{0.1cm}}c|@{\hspace*{0.1cm}}c}
\hline\\[-3pt]
system&Ref.&$L_{max}$&$\alpha$&$\beta$&$\gamma$&$\nu$&$\eta$\\[2pt]
\hline\\[-8pt]
STAR\index{STAR}&\cite{LoisonSchotteO4}&42&0.287(27)$^{(b)}$&0.291(11)&1.133(28)&0.571(9)&+0.015(18)$^{(a)}$\\[2pt]
\hline\\[-8pt]
$V_{4,2}$\index{dihedral lattice} &\cite{LoisonSchotteO4}&40&0.278(30)$^{(b)}$&0.290(12)&1.142(34)&0.574(10)&+0.011(25)$^{(a)}$\\[2pt]
\hline
\end{tabular}}
\label{table_O4}
\end{table}

\begin{table}
\tbl{ Critical exponents \index{critical exponents} by Monte Carlo
for spins with six components ($N=6$) and a BS $O(6)/O(4)\equiv
SO(6)/SO(4)$. Calculated by $^{(a)}\;\gamma/\nu=2-\eta\,$,
$^{(b)}\;d \nu=2-\alpha\,$. } {\tabcolsep8pt
\begin{tabular}{@{}c|@{\hspace*{0.1cm}}c|@{\hspace*{0.1cm}}c|@{\hspace*{0.1cm}}c|@{\hspace*{0.1cm}}c|@{\hspace*{0.1cm}}c|@{\hspace*{0.1cm}}c|@{\hspace*{0.1cm}}c}
\hline\\[-3pt]
system&Ref.&$L_{max}$&$\alpha$&$\beta$&$\gamma$&$\nu$&$\eta$\\[2pt]
\hline\\[-8pt]
STA\index{STA}&\cite{Loison_O6}&36&-0.100(33)$^{(b)}$&0.359(14)&1.383(36)&0.700(11)&+0.025(20)$^{(a)}$\\[2pt]
\hline
\end{tabular}}
\label{table_O6}
\end{table}

\subsubsection{Experiment}

A resembling situation occurs in the analysis of experiments.
There the correlation length $\xi$ plays the role of the system
size in numerical simulations. For a second-order transition one
has $\xi\sim (T-T_c)^{-\nu}$. Therefore we should observe a
crossover between the region of low velocity (``almost
second-order transition'' in Zumbach's words) to the true
first-order behavior for temperatures ``closer'' to the critical
temperature. Unfortunately the situation is even more complicated
in experimental systems with the omnipresence of planar or axial
anisotropies and the one- or two-dimensional characters of the
compounds. Then  a succession of crossovers (see \ref{appendixRG}
for more details about crossovers) from 2d to 3d and from
Heisenberg to Ising or $XY$ behavior could lead to difficulties in
the interpretation. Furthermore other (small) interactions could
also dominate the behavior near the critical temperature and
change the universality class.

\vspace{\baselineskip} {\bf Experiments for $XY$ spins\index{XY
spin}}: All experimental results can be found in Ref. [34].
Several $AXB_3$ compounds have the STA structure. The experiments
on CsMnBr$_3$
,\cite{Mason2,Mason3,Ajiro,Gaulin,Mason1,Kadowaki1,Wang1,Weber,Goto,Deutschmann,Collins1}
RbMnBr$_3$\cite{Kato1,Kato2} and CsVBr$_3$\cite{Tanaka} give
critical exponents compatible with those of MC simulation on STA
and a second-order transition. We can interpret this result by the
fact that the systems are under the influence of a complex fixed
point, and $t\propto T-T_c$ is too small to observe a first-order
transition.

The case CsCuCl$_3$\cite{Weber2} is different since the authors
observe a crossover from a second-order region with exponents
compatible with MC results on STA for $10^{-3} < t < 5.10^{-2}$ to
a region of first-order transition for $5.10^{-5} < t <
5.10^{-3}$. For $t < t_0 \approx 10^{-3}$ one seems to observe the
true first-order region which corroborates the scenario introduced
previously.

Three kinds of helimagnetic structure have been studied: Holmium,
Dysprosium and Terbium. Essentially three types of results exist:
those compatible with MC ones of the STA, those with a large
$\beta$ incompatible with STA and those showing a weak first-order
transition.

The results compatible with those of MC on STA for Ho
\cite{Jayasuriya1,Thurston1,Gaulin1}
Dy\cite{Gaulin1,Lederman,Jayasuriya} and
Tb\cite{Jayasuriya3,Dietrich,Tang1,Hirota,Tang2} can be
interpreted as before: the systems are under the influence of
$F_+$. The first-order transition for Ho\cite{Tindall1} and
Dy\cite{Zochowski,Astrom} is due to the fact that the measurements
were done in the first-order region near the transition
temperature. The values of the exponent $\beta\sim0.39$ in the
case of Ho\cite{Thurston1,DuPlessis1,Eckert1,Helgesen} and
Dy\cite{DuPlessis1,DuPlessis2,Brits,Loh} are not compatible with
those found by MC ($\beta\sim0.25$). This fact can be explained by
the presence of a second length scale in the critical fluctuations
near $T_c$ related to random strain fields which are localized at
or near the sample surface.\cite{Thurston1} Thus the critical
exponent $\beta$ measured depends on this second length.

\vspace{\baselineskip} {\bf Experiments for
Heisenberg\index{Heisenberg spin}  spins}: All experimental
results can be found in Ref. [34]. As explained previously, before
the first-order region is reached, the crossover from Heisenberg
to Ising or $XY$ behavior prevents a first-order transition of
Heisenberg type. Nevertheless the second-order transition can be
studied for VCl$_2$,\cite{Kadowaki} VBr$_2$\cite{Wosnitza},
Cu(HCOO)$_2$2CO(ND$_2$)$_2$2D$_2$O\cite{Koyama} and
Fe[S$_2$CN(C$_2$H$_5$)$_2$]$_2$Cl.\cite{DeFotis} For the last two
examples the observed exponents might be influenced by the
crossover from 2d to 3d Ising behavior. The experimental results
agree quite well with the MC simulations.

\subsubsection{Value of $N_c$}

Tables~\ref{table_O2}-\ref{table_O6} allow us to get an idea of
the value of $N_c$. The MCRG gives an estimate $3<N_c<8$ but the
values of $\eta$ can give a better estimate following Ref. [34].
As can be seen for the $XY$ and Heisenberg systems, negative
values of $\eta$ appear for the STA, STAR or dihedral model. But
$\eta$ must always be positive.\cite{Patashinskii,Zinn89} This is
due to the use of the scaling relation $\gamma/\nu=2-\eta$. Indeed
Zumbach\cite{Zumbach93} has shown that for an ``almost
second-order transition'', i.e. when the solution becomes complex,
this relation has to be modified to $\gamma/\nu=2-\eta+c$, $c$
being a constant different from zero. We can use this relation as
a criterion for real or complex fixed points. In three dimensions
$\eta$ is usually small and is almost independent of $N$ for the
ferromagnetic case, that is $\sim 0.03$ (see
Table~\ref{table_ferro}). Our hypothesis is that it is also true
for the frustrated case. Indeed for $N=6$ we found $\eta\sim 0.03$
(see Table~\ref{table_O6}). Accepting this value $c$ becomes zero
around $N_c\sim 4.5$. Obviously if a bigger value of $\eta$ is
chosen, $N_c$ will increase.

\subsubsection{Phase diagram (N,d)}

\begin{figure}[th]
\centerline{\psfig{file=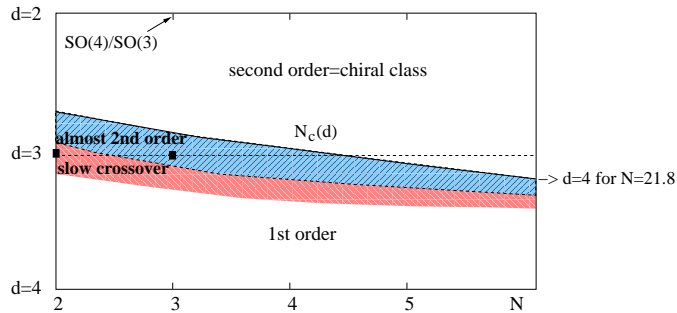,width=3.5in,angle=0}}
\vspace*{8pt} \caption{ The critical curve $N_c(d)$ separating the
first-order region for small $N$ and large $d$ from the
second-order region for large $N$ and low $d$. The squares
represent the physical systems of interest ($N=2,3$ in three
dimensions). \label{fig_Nc_D} }
\end{figure}

In Fig.~\ref{fig_Nc_D} we have plotted a phase diagram where the
abscissa is the spin dimension $N$ and the ordinate the space
dimension $d$.

There is a line which divides a region of first-order transition
from a region of second-order transition. At, and near, $d=4$ we
can use the RG $d=4-\epsilon$ expansion to find
$N_c=21.8-23.4\,\epsilon$, and for $d=3$ we have seen that
$N_c\sim 4.5$. Furthermore there is a region near this line where
the transition is ``almost second order'', i.e. under the
influence of the complex fixed point for ``small'' sizes. The
system will display a second-order transition with stable critical
exponents. Besides, there is a slow crossover region where
critical exponents vary with the system size. The sizes of these
regions depend on the initial point in the flow diagram and
therefore on the model.

Near two dimensions and for Heisenberg spins the transition is of
the type $SO(4)/SO(3)$, i.e. ferromagnetic for spins with four
components (see the following section and the chapter of Delamotte
et al.\cite{chapter_delamotte} for more details).

\subsubsection{Renormalization-Group expansions}

We present here a review of the RG expansions (``non
perturbative'', $d=4-\epsilon$, in fixed dimension $d=3$, and
$d=2+\epsilon$). For more details see \ref{appendixRG} and the
chapter of this book of Delamotte et al.\cite{chapter_delamotte}

We discuss now the $4-\epsilon$ expansion. Using a continuous
limit of the Landau Ginzburg Hamiltonian
(\ref{hamiltonian_uns_dihe}), it involves a perturbative extension
($u,v$) around the Gaussian solution with the dimension $d$ and
the number of spin components $N$ as parameters. To get the result
for $d=3$ one puts $\epsilon=1$. Of course the series are at best
asymptotic and must be resummed, even for the ferromagnetic case
.\cite{LeGuillou} The extension until $\epsilon^2$ in Ref. [88]
has been ``resummed'' (only three terms) and the value of
$N_c(d=3)\sim 3.39$ is compatible with the numerical result
$N_c^{numerical}(d=3)\sim 4.5$. The calculations have been
extended to $\epsilon^4$ in Ref. [89] but unfortunately no
resummation has been done on this series yet.

The expansion for fixed dimension $d=3$ is very similar to the
$d=4-\epsilon$ one. The series and the resummation will differ
slightly. At three loops the result is $N_c\sim
3.91$\cite{Antonenko 94} which is again not far from
$N_c^{numerical}(d=3)\sim 4.5$. A very interesting picture emerges
from the six-loop calculations.\cite{Pelisseto,Calabrese} First,
for $N\sim 5$, the resummation does not converge. Second, for
$N=3$, the resummation gives rise to a very large variation of the
critical exponents following the flow chosen to arrive at a single
critical exponent at the critical point (see Fig. 5 of the
reference\cite{Calabrese}) which is a really ``new'' phenomenon.
Third, for $N=2$ a second-order transition is predicted in
contradiction to Itakura's results\cite{Itakura2001} and the
numerical simulations (see Tables~\ref{table_O2}-\ref{table_O2K}).
Manifestly we could conclude that the resummation chosen does not
work for this series. To understand this we have to stress that
the series has many more terms because of the double expansions in
$u$ and $v$ compared to a single one for the ferromagnetic case.
Whence it is much more difficult to find a ``good'' resummation
scheme.

On the other hand Delamotte et al.\cite{Delamotte2} found that the
transition must be of $SO(4)/SO(3)$ type, i.e. equivalent to a
ferromagnetic $O(4)$ transition, near two dimension for Heisenberg
spins. This study was done using the $d=2+\epsilon$ expansion or
equivalently the non linear $\sigma$ model and the result is valid
only near two dimensions. This prediction was checked by Southern
and Young\cite{Southern_Young} by MC in two dimensions. But we can
give arguments\cite{LoisonSchotteHei} using $1/N$--expansion and
the value of the critical exponents to rule out this kind of
transition in three dimensions. These arguments hold as long as
the relevant operators are identical in $1/N$ expansion and in the
non linear $\sigma$ model. Indeed for the $1/N$ expansion, and
equivalently for the $4-\epsilon$ expansion and in fixed
dimension, we keep in the Hamiltonian only the terms which are
renormalizable near $d=4$ or infinite $N$, and discard the others.
Nevertheless it seems that some non renormalizable terms (see
\ref{appendixRG}) become relevant and important for low $N$ and
$d$ and, whatever the number of loops we cannot find the correct
behavior using these expansions.

Because of  the problems encountered by the standard expansions, a
``non perturbative'' approach could be very useful. This was
introduced in this model by Tissier et al.\cite{Tissier} We quote
``non perturbative'' because it is not an expansion resembling the
other methods where the extension parameters ($u,v,\epsilon$) are
usually not small. In this non perturbative method even if we
introduce only a few terms in the action,  results will be very
good and no resummation is necessary. We notice that the simplest
action was studied by Zumbach\cite{Zumbach93} which allows him to
introduce the ``almost second-order'' transition. Adding more
terms, Tissier at al. found that they are able to retrieve all the
previous expansions with additional information. Their value of
$N_c\sim 5.1$ is comparable to $N_c^{numerical}\sim 4.5$. In three
dimensions they found critical exponents very close to those
calculated by MC for $N=6$ (see Table~\ref{table_O6}),  a minimum
in the flow for $N=3$, and a slow crossover for $N=2$ as explained
in the previous section. They found also that the critical
behavior is indeed that of a ferromagnetic $O(4)$ transition near
two dimensions for Heisenberg spins. Some non--renormalizable
operators excluded in the $4-\epsilon$ and $1/N$ expansions are
included in this ``non perturbative'' method. These operators are
always relevant between two and four dimensions whatever $N$ is,
but have an influence on the values of the critical exponents only
for small $N$ and near two dimensions. This explains the
discrepancy between the $d=2+\epsilon$ and $d=4-\epsilon$
expansions.

\subsubsection{Short historical review}
In this section we give a short historical review of studies on
frustrated systems. Indeed the history was not straight if we look
back at the last 25 years. The development began by a RG
$4-\epsilon$ expansion by Jones, Love and Moore\cite{Jones} in
1978. It is only in 1984 and in the following years that
Kawamura\cite{Kawa92} started with the first numerical simulations
and, in combination with the results of several experiments,
proposed a ``new'' universality class.

Then Azaria, Delamotte et al\cite{Delamotte2} found that the
transition near two dimensions for Heisenberg spins should belong
to the $O(4)$ ferromagnetic class. Several groups (see
Tables~\ref{table_O2}-\ref{table_O6}) have done numerical
simulations and experiments which favored either the new
universality class, a first-order transition, an $O(4)$ transition
and even a mean-field tricritical transition. During those years
Sokolov et al.\cite{Antonenko2,Antonenko 94} have extended the RG
expansion to three loops and found a first-order behavior for $XY$
and Heisenberg spins.

Surely one of the most important articles to find the key to
understand the physics was written by Zumbach,\cite{Zumbach93}
using a non perturbative approach. He predicts an ``almost
second-order transition'' for $XY$ and Heisenberg spins. Loison
and Schotte,\cite{LoisonSchotteXY,LoisonSchotteHei} using this
concept, were able to get a clear picture for both the numerical
and experimental studies. Then Tissier, Delamotte and
Mouhanna,\cite{Tissier}, by extending the work of Zumbach, were
able to understand the whole phase diagram for the dimension
between two and four. To terminate Itakura,\cite{Itakura2001}
confirmed the picture given by Zumbach\&Loison\&Schotte, which we
think is the definitive answer from a numerical point of view.

\subsubsection{Relations with the Potts\index{Potts model}  model}
\begin{figure}[th]
\centerline{\psfig{file=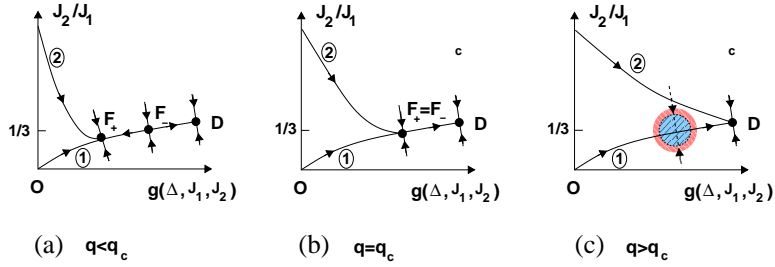,width=4.0in,angle=0}}
\vspace*{8pt} \caption{ Hypothesis of the renormalization flow in
presence of second-neighbor interaction. For $q \le q_c$ the
initial point of the flow is irrelevant to the critical behavior.
For $q>q_c$ the behavior for accessible lattice size will be
different for the curve~1 (going through the basin of attraction
of the fixed points $F_+$) and the curve~2 (going directly to the
first-order fixed point $D$). \label{fig_Potts} }
\end{figure}

We would like now to stress the similarities between the Potts
model and the frustrated case just studied.\cite{Loison_potts} In
Fig.~\ref{fig_Potts} we have plotted the RG flow
diagram\cite{Nienhuis} as a function of the first- and
second-neighbor ferromagnetic interactions ($J_1$ and $J_2$) and
the chemical potential $\Delta$ (corresponding to the site
vacancy).

$F_+$ is the standard stable ferromagnetic fixed point and $F_-$
is an unstable fixed point. $D$ is a first-order fixed point. $O$
is the initial point of the system. This figure resembles the one
of the frustrated case (Fig.~\ref{fig_RG_tri}). For $q<q_c$ the
flow goes to the fixed point $F_+$ and has a second-order
transition. At $q=q_c=4$ (in two dimensions) the two points $F_+$
and $F_-$ collapse and a second-order phase transition with
logarithm correction results. For $q>q_c$ there is no longer a
solution and the flow goes directly to the fixed point $D$, i.e.
the system has a first-order transition. Our hypothesis is that
even if the solution of $F_+$ is complex, it has an influence on
the real space and there is a region of low velocity in the flow
diagram. It can be a true minimum for $q\gtrsim q_c$ with an
``almost'' second-order transition. For $q$ bigger, no minimum
should exist but, following the sizes studied, a variation of the
critical exponents could occur. Indeed for $q=5$  a set of
critical exponents was found,\cite{Peczak_Potts} corresponding to
our hypothesis of ``almost second-order'' transition. The cases
$q=6$ and $q=7$ have not been studied, but possibly one could
observe a variation of critical exponents corresponding to a slow
crossover to the first-order point.

We remark that  depending of $q$ we can observe clearly if the
transition is of first order or second order. In
Fig.~\ref{fig_Potts} we have plotted the flow (2) in presence of
the second-neighbor interaction. We observe that this flow is less
influenced by the low velocity region than the flow (1) if $q>q_c$
and it reaches $F_+$ if $q \leq q_c$. Adding the
third-nearest-neighbor interaction, we are able to find
numerically that $q_c=4$ for the crossover from the first-order to
second-order transitions. Contrary to the frustrated case we know
roughly the coordinate of the fixed point $F_+$ ($J_2/J_1\sim
0.3$)\cite{Nauenberg} and $\Delta\neq 0$.\cite{Nienhuis} By
consequence we know which kind of interaction must be added.

\subsection{$O(N)/O(N-P)$ breakdown of symmetry for $d=3$}

In mean-field theory, for $N> P$ the model shows a usual
second-order type, but for $N=P$ the transition is a special
one.\cite{Zumbach93} The BS in this case is $Z_2 \otimes SO(N)$
and the coupling between the two symmetries leads to a special
behavior which, for $N=P=2$ in two dimensions, has been
extensively debated (see later in this chapter).

These generalized chiral models have been studied by applying the
RG  technique ($d=4-\epsilon$
expansion).\cite{Zumbach93,Saul,Kawamura90} The picture is very
similar for all $N \ge P \ge 2$. At the lowest order in
$\epsilon$, there are up to four fixed points, depending on the
values of $N$ and $P$. Amongst them are the trivial Gaussian fixed
point and the standard isotropic $O(NP)$ Heisenberg fixed point.
These two fixed points are unstable. In addition, a pair of new
fixed points, one stable and the other unstable, appear when $N\ge
N_c(d)$ with
\begin{eqnarray}
N_c(d)&=&5P+2+2\sqrt{6(P+2)(P-1)} \notag \\
&&-\bigg{[}5P+2+\frac{25P^2+22P-32}{2\sqrt{6(P+2)(P-1)}}\bigg{]}\,\epsilon \,.
\end{eqnarray}
For $P=2$ we find the standard result $N_c=21.8-23.4\,\epsilon$.
On the other hand, for $P=3$ we obtain $N_c=32.5-33.7\,\epsilon$
and for $P=4$ we obtain $N_c=42.8-43.9\,\epsilon$. A "tricritical"
line exists which separates a second-order region for low $d$ and
large $N$ from a first-order region for large $d$ and small $N$.
Applying $\epsilon=1$ ($d=3$), we obtained that $N_c(d=3)<0$ for
all $P$. For $P=2$ we know that this result does not hold and
equivalently it does not apply for $P\ge 3$.
Loison\cite{Loison_V_NP} has done some simulations for $N=3$ and
$N=4$ with $P=N$ and $P=N-1$ for the Stiefel\index{Stiefel model}
$V_{N,P}$ model. He showed that the transition is clearly of first
order using numerical simulations. He generalized this result for
all $N$. We remark that the fully frustrated cubic lattice
(generalized Villain lattice) which could have an identical BS has
also a first-order transition for Heisenberg spins.\cite{Alonso96}
In conclusion the transition in systems with a $O(N)/O(N-P)$ the
breakdown of symmetry is of first order for $P=N$ and $P=N-1$ and
in particular for the physical case $N=P=3$, i.e. a non planar GS
for Heisenberg spins.

\subsection{$Z_2 \otimes SO(N)/SO(N-1)$ breakdown of symmetry for $d=3$}

For $N>2$, this BS corresponds to a collinear ground state
associated to a BS of the lattice (see
Table~\ref{table_BS_O2}-\ref{table_BS_O3}). For example the
stacked Villain lattices with first-neighbor ferromagnetic
interaction $J_1$ and a second-neighbor antiferromagnetic
interaction $J_2$ can have this BS. For $J_2/J_1<-0.5$ the
competition between the two interactions leads to two possible
ground states.\cite{LoisonSimon} For $N=2$ ($XY$ spins) the
stacked Villain lattices and the stacked $J_1$--$J_2$ model have
the same BS as the STA and should display an analogous behavior
(``almost second order'' for small sizes following by a
first-order transition for larger sizes).

We have simulated by MC technique the stacked
Villain\index{Villain lattice} lattices and the stacked
$J_1$--$J_2$ lattice which have this kind of BS. We found that the
transition seems continuous\cite{Loison_preparation} for small
lattices for $XY$ and Heisenberg spins. For $XY$ spins this is in
agreement with the numerical simulation of STA which has an
identical BS. We have also studied the Ising--$V_{N,1}$ model
which has the same BS. For $N=2$ this model is the equivalent to
the $V_{2,2}$ model which has a strong first-order behavior as
explained previously. For $N=3$ and $N=4$ we found also strong
first-order transitions. Therefore it seems that the transition
$Z_2 \otimes SO(N)/SO(N-1)$ is of first order for any $N$ even if
it could exist an ``almost second-order transition'' for small
sizes.

\subsection{$Z_3 \otimes SO(N)/SO(N-1)$ breakdown of symmetry for $d=3$}

This BS appears with a collinear ground state associated to a
three-state Potts symmetry due to breakdown of lattice symmetries.
This corresponds to various lattices where the unit cell is
composed of four spins (two parallel spins and two antiparallel
spins) like the STA with intermediate NNN interaction
($0.125<J_2/J_1<1$), the fcc, the hcp, the cubic $J_1$--$J_2$
model or the pyrochlore (see
Table~\ref{table_BS_O2}-\ref{table_BS_O3}).

As seen previously, even the coupling of an Ising symmetry ($Z_2$)
with $SO(N)/SO(N-1)$ can give a first-order transition. Whence it
is expected that the transition with a three-state Potts symmetry
($Z_3$) gives also a first-order transition which is even stronger
if we consider that the Potts symmetry alone has a first-order
transition in $d=3$ for $q=3$. Indeed a strong first-order
transition was seen in the fcc,\cite{Alonso96,Diep_fcc}
hcp,\cite{Diep_hcp} cubic $J_1$--$J_2$
model\cite{Alonso96,Diep_cubic_J1_J2} pyrochlore\cite{Reimers2}
and STA with NNN interaction\cite{Loison 96,LoisonHei} for $XY$
and Heisenberg spins.

We have done some simulations for the Potts--$V_{N,1}$ model with
the Hamiltonian (\ref{H_ON_Potts}) with $q\geq 3$, i.e. the
$q$-state Potts model coupled to an $N$-component vector. We found
a strong first-order transition for any $N$ and $q$.

In conclusion the transition for the $Z_q \otimes SO(N)/SO(N-1)$
BS in $d=3$ is of first order for any $N\geq 2$ and $q\geq 2$.

\subsection{$Z_q \otimes O(N)/O(N-2)$ and other breakdown of symmetry in $d=3$}

This BS appears with a planar ground state associated to a
three-state Potts symmetry. One example is the STA with NNN
interaction $J_2/J_1>1$ (see
Table~\ref{table_BS_O2}-\ref{table_BS_O3}).

To simulate this BS we have used a Potts--$V_{N,2}$ model for any
$q$. We found the behavior of a strong first-order transition
whatever $N$ and $q$ are. It is not surprising if we consider that
the ``less frustrated'' model Potts--$V_{N,1}$ is already of first
order.

We have also simulated the $Z_q \otimes O(N)/O(N-3)$  for $N=3$
which corresponds to the BS $Z_q \otimes Z_2 \otimes SO(N) $ using
a Potts--$V_{N,P}$ model. We found again a strong first-order
transition. Similar results are obtained for
Ising--Potts--$V_{N,1}$ and  Ising--Potts--$V_{N,2}$ models.

To summarize, it seems that the coupling between a Potts or Ising
symmetry and a continuous symmetry of vector spins give always
rise to a first-order transition.

\section{Conclusion}

We have studied breakdowns of symmetry of three-dimensional
lattices for $XY$ ($N=2$) and Heisenberg ($N=3$) spins. The
general form of the breakdown of symmetry is $S_{lattice} \otimes
O(N)/O(N-P)$ . $S_{lattice}$ is usually a Potts symmetry $Z_q$
($Z_1$ corresponds to the identity, $Z_2$ to an Ising symmetry,
$Z_3$ to a three-state Potts symmetry, $\dots$). $P$ runs from 1
to 3.

If the frustration is small the GS can be collinear ($P=1$)
without breaking any symmetry of the lattice $S_{lattice}=1$, the
BS will be  $ SO(N)/SO(N-1)$ and the transition will be of second
order like in ferromagnetic systems.

If the frustration is strong enough, we can have a non collinear
GS with the usual planar configuration ($P=2$) like in the STA or
helimagnets and $S_{lattice}=1$. Depending on the model and for
not ``too big'' sizes, the transition could appear continuous
belonging to a new universality class. For an infinite size the
system will have a first-order transition.

Some more complicated systems could also have a non planar GS, and
for Heisenberg spins it corresponds to $P=3$ with usually
$S_{lattice}=1$. The transition will be of first order with a
possible ``almost second-order'' behavior for small system sizes
similar to the $P=2$ case.

Some even more complicated systems could have a non collinear GS
($P=2$ or $P=3$) with $S_{lattice}=Z_q$ and $q=3$. For example the
STA with large NNN interaction has a  strong first-order
transition.

On the other hand some frustrated systems have degenerate GS but
by ``order by disorder''  a collinear GS ($P=1$) is selected. The
lattice symmetry could be broken giving an additional BS.
$S_{lattice}=Z_2$ (Ising) for the stacked Villain model, or
$S_{lattice}=Z_3$ in fcc, hcp, STA with intermediate NNN
interaction, and pyrochlore. For the $Z_3$  symmetry the
transition is strongly of first order even for small sizes. For
the $Z_2$ symmetry the transition is also of first order for large
sizes but looks continuous for small sizes. Therefore we cannot
exclude an ``almost second-order'' behavior belonging to a new
universality class.

In conclusion frustrated systems have a first-order transition for
$XY$ and Heisenberg spins in three dimensions even if for
``small'' sizes the systems would show a second-order transition.
``Small'' could mean sizes of thousands of lattice constants and
the first-order transition will then not be observable with the
actual computer resources.

\section{$O(N)$ frustrated vector spins in $d=2$}

\subsection{Introduction}

The situations in two-dimensional systems are different from the
three dimensions. The Mermin--Wagner theorem\cite{Mermin66}
asserts that no magnetization appears at non-zero temperature.
However, a transition due to the topological
defects\cite{Mermin_defects,Toulouse_defect} can appear. The
binding--unbinding transition of vortex--antivortex pairs for $XY$
spin systems is a classical example. We note that their exact role
for three-dimensional phase transitions is not
clear.\cite{Kohring,Williams,Shenoy} For non collinear GS, the
topological properties of the system differ from those of
collinear GS and new types of transition could appear. In the
following we will briefly review the ferromagnetic case, then the
frustrated $XY$ case, applicable to Josephson--junction arrays in
magnetic fields, and finally the frustrated Heisenberg cases.

\subsection{Non frustrated $XY$ spin systems\index{XY spin}  }

For a non frustrated $XY$ spin system the order parameter is
$SO(2)$. The topological defects of this group is only the point
defect classified by the homotopy group $\Pi_1(SO(2))=Z$, $Z$
being the topological quantum number of the
defect.\cite{Mermin_defects,Toulouse_defect} A
Kosterlitz--Thouless (KT) transition \cite{Thouless73} exists at
the critical temperature $T_c\sim0.9$, driven by the unbounding of
vortex--antivortex pairs.

This transition has some special features: for $T<T_c$ the
correlation length ($\xi$) is infinite, while for $T>T_c$ it has
an exponential decreasing contrary to the power-law behavior in
the standard transition.

Several numerical methods exist to calculate the critical
temperature and the critical exponents (see \ref{appendixMC}).

\subsection{Frustrated $XY$ spin systems: $Z_2\otimes SO(2)$ \label{2d_XY_Z2}}
A frustrated $XY$ spin system can have many order parameters as
previously seen (see Table~\ref{table_BS_O2}). The most studied
case is $Z_2\otimes SO(2)$ which corresponds to a non collinear
GS: triangular\index{triangular lattice}
lattice,\cite{JLee91,Miyashita84,LeeLee98,Caprioti97,Loison_TA_XY}
Villain\index{Villain lattice}  lattice or two-dimensional fully
\index{fully frustrated lattice} frustrated
lattice\cite{JLee91,vill1,Teitel83,Berge86,Nicolaides91,Jose96,Thijssen90,Granato93,LeeLee94,Luo98}
and related Zig--Zag models.\cite{Boubcheur98,Benakly97} In
combination to the topological defect\index{topological defect}
point $Z$, there is a line of $Z_2$ corresponding to an Ising
transition. Therefore they will have similar properties as the
$J_1$--$J_2$ model, the $V_{2,2}$ model or any model where the
$Z_2$ symmetry comes from the lattice. We note that different
models can have an identical group and therefore the same
topological defects. The following models should have analogous
properties: the Villain model on the Villain
lattice,\cite{Olsson95} The Ising--$V_{2,1}\equiv V_{2,2}$
model,\cite{Nightingale95,Granato91} the 19-vertex
model,\cite{Knops94} the $1D$ quantum spins,\cite{Granato92} the
Coulomb gas
representation,\cite{Yosephin85,Minnhagen85,Grest89,JRLee94} the
$XY-XY$ model,\cite{Choi85,Jeon97} or the RSOS
model.\cite{LeeLeeRSOS}

This kind of models was widely studied because it is believed to
correspond to experimental systems such as  Josephson--Junction
arrays of weakly coupled superconducting
islands\cite{Lobb88,Ling96} or films of Helium
$^3$He.\cite{Halsey85,Korshunov86,Kotsubo87,Xu90}

The question is to understand the coupling between the
vortex--antivortex and the Ising symmetry. Contrary to the
three-dimensional case discussed previously there are no clear
accepted answers to these questions. We will therefore  only
present the different results in the high temperature region and
in the FSS, giving some clues to understand the physics of theses
systems.

What are the possible scenarios?
\begin{itemize}
\item[1.] $T_c^{KT}<T_c^{Ising}$\\
The KT transition\index{Kosterlitz-Thouless transition} associated
to the topological defects appears at  a temperature lower than
the one associated to the Ising transition even if they can be
very close. The two transitions should
have standard behaviors ($\nu^{KT}=0.5$, $\nu^{Ising}=1$, $\cdots$).\\
Furthermore Olsson has proposed that the Ising behavior could be
observed only when  $L\gg \xi^{KT}$. Since $\xi^{KT}$ is infinite
for $T\leq T_c^{KT}$ and decreases exponentially for $T>T_c$, it
will correspond to a high temperature region with $T\gg T_c^{KT}$,
but also $T\gg T_c^{Ising}$ when the two transitions appear in
close neighborhood. In particular the FSS region will have a non
standard Ising behavior for the sizes accessible with actual
computer resources. Only for a very large lattice the standard
Ising behavior will be visible. We observe the similarity of this
hypothesis with the one described previously in three dimensions
and we could call the transition in the FSS an ``almost new
Ising--KT behavior''.

\item[2.] $T_c^{KT}\sim T_c^{Ising}$\\
The two transitions appear at the same temperature and display
different behaviors compared to the standard one. We remark that
for high temperatures $T\gg T_c$, i.e. in the high-temperature
region, we cannot assume that the coupling between the $Z_2$
symmetry and the topological defects would be identical as for
$T\sim T_c$. Indeed with a lot of small $Z_2$ walls dividing the
system it is not certain that the vortex and the antivortex should
play an important role and we could get a standard Ising
transition.

\item[3.] $T_c^{KT}>T_c^{Ising}$\\
This hypothesis was advanced by Garel et Doniach in 1980 for the
$J_1$--$J_2$ model,\cite{Garel} but numerical MC simulations have
shown that it is not the case\cite{LoisonSimon} for this system.
\end{itemize}

To choose the most probable hypothesis we will look now at the
numerical results:

Some information can be provided by numerical MC simulations in
the high-temperature region. Fitting the results it is possible to
get the critical temperature and the critical exponents. Because
of the presence of corrections and by consequence of many free
parameters, the extraction of the results is difficult.

For the Ising symmetry Olsson\cite{Olsson95} has found a standard
ferromagnetic one with $\nu^{Ising}\sim 1$) while Jose and
Ramirez\cite{Jose96} found $\nu^{Ising}\sim0.87$. New simulations
should be done to resolve this discrepancy. If $\nu^{Ising} \neq
1$, it will a counter--proof to Olsson's claim (first
possibility). On the other hand, if $\nu^{Ising}=1$, we cannot
rule out the second possibility.

The exponent of the KT transition have been only calculated  by
Jose and Ramirez.\cite{Jose96} They found non standard
$\nu^{KT}\sim 0.3$ to be compared to 0.5 in the ferromagnetic
case. If this result is in favor of the second hypothesis, it
cannot rule out the first one. Indeed we can reverse the argument
given at the end of the second hypothesis. If at high temperatures
the system is composed of many domains separated by walls of
$Z_2$, the correlation between the topological defects could not
be the one of the standard one. Therefore there is no definitive
conclusion from the high-temperature simulations.

Now we can look at the results in the FSS.
\\
For the Ising transition the critical exponents have been
precisely determined by Loison and
Simon,\cite{LoisonSimon,Loison_TA_XY} for the $J_1$--$J_2$ model
using\index{J1-J2 model}\index{square lattice}  both the FSS and
the dynamical properties  including the first
correction\cite{Loison_TA_XY} due to the lattice size $L$. They
obtained non standard critical exponents $\nu^{Ising}=0.815(20)$.
The ``large'' error is due to the inclusion of the corrections.
See Ref. [111] for a review of all the numerical works. For the
triangular\index{triangular lattice} lattice the critical
temperature is known with high precision $T_c^{Ising}=0.5122(1)$.
\\
The KT\index{Kosterlitz-Thouless transition}  transition is more
problematic. First we can use the helicity\cite{Thouless73}
$\Upsilon$ to find the critical temperature. If we admit the same
universal jump at the critical temperature as the ferromagnetic
one, we get $T_c^{KT}=0.5010(10)$ a temperature much smaller than
$T_c^{Ising}=0.5122(1)$. But we cannot be sure that
$\Upsilon_{jump}^{frustrated}=\Upsilon_{jump}^{ferromagnetic}$
and this result must be taken with care.\\
As explained in the previous section we can use the Binder
parameter or the dynamical properties of the model to calculate
$T_c^{KT}$ and the critical exponents. Using this method we found
very interesting results. First the Binder parameter shows a
power-law transition and not an exponential one. In addition the
critical\index{critical exponents} exponents and temperatures
calculated by the two methods are in good agreement with
$T_c^{KT}=0.5102(2)$ just below $T_c^{Ising}=0.5122(1)$ for the
triangular lattice. The critical exponent $\eta^{KT}=0.36(1)$ is
very different from the standard one 0.25. These results are in
accord with those on the $J_1$--$J_2$ model. What is disturbing is
that for the triangular lattice, accepting $T_c^{KT}=0.5102(2)$,
the helicity\index{helicity}  jump is smaller than the
$\Upsilon_{jump}^{ferromagnetic}$
although it is believed that it should be greater.\\
The situation is puzzling with no definitive conclusions.
%

\subsection{Frustrated $XY$ spin systems: $Z_3\otimes SO(2)$}
Two-dimensional systems with an  order parameter like $Z_3\otimes
SO(2)$ are less numerous than three-dimensional systems. Following
Table~\ref{table_BS_O2} we know at least two systems: the
triangular\index{triangular lattice}  lattice with intermediate
NNN interaction and the Potts--$V_{2,1}$ model.

No  numerical studies have been done for the former system, but we
have studied\cite{Loison_XY_Z3} the latter one. It gives a
first-order transition.

Following the reasoning of the previous section two principal
hypotheses appear:
\begin{itemize}
\item[1.] Olsson hypothesis: KT transition at lower temperature
following the transition associated to the Potts symmetry $Z_3$ at
higher temperature. Non standard critical exponents should appear
in the FSS region due to a screening length ($\xi^{KT}$). The
``true'' standard second-order behavior will appear only for very
large lattices ($L\gg \xi^{KT}$) or in the high-temperature
region. \item[2.] A new behavior for the KT and Potts transitions
at the same critical temperature.
\end{itemize}

The results are in favor of the second hypothesis. The Potts
symmetry has a first-order transition. It means that its
correlation length $\xi_{Z_3}$ is finite at the critical
temperature and less than $\xi_{KT}$. There will be no change even
if the system size $L$ is much larger and the system will never
show a standard three-state Potts second-order transition.
Obviously we cannot apply directly this result to the Ising--KT
transition since the two models are not equivalent although it is
nevertheless an argument against the Olsson's hypothesis.

\subsection{Frustrated $XY$ spin systems: $Z_2 \otimes Z_2 \otimes SO(2)$ and $Z_3 \otimes Z_2 \otimes SO(2)$}
Following Table~\ref{table_BS_O2} the order parameter $Z_2\otimes
Z_2 \otimes SO(2)$ can appear with a $J_1$--$J_2$--$J_3$ lattice
or an Ising-$V_{2,2}$ model. In addition $Z_3\otimes Z_2 \otimes
SO(2)$ exists in the triangular antiferromagnetic with NNN
interaction and in the Potts--$V_{2,2}$ model. No numerical
studies have been done on these models.

\subsection{Frustrated Heisenberg  spin systems: $SO(3)$ }\index{Heisenberg spin}
For a non frustrated Heisenberg spin system the order parameter
will be $SO(3)/SO(2)$ and no topological\index{topological defect}
defects exist in two dimensions, no phase transition will appear.

For a frustrated system many order parameters can exist (see
Table~\ref{table_BS_O3}), but only the planar GS have received
attention. In this case the order parameter is $SO(3)$ and there
exist point defects that is
$Z_2$--vortex--antivortex\cite{Mermin_defects,Toulouse_defect}.
These topological defects are different from the $Z$--vortex
present in $XY$ systems. The existence of a critical transition
driven by the unbounding of vortex--antivortex was first
conjectured by Kawamura and Miyashita.\cite{Kawamura_Miyashita}

At low temperatures the spin waves will dominate the behavior of
the system and forbid an infinite correlation length below $T_c$
contrary to  $XY$ spins. The behavior should be equivalent to the
one present in the four-dimensional ferromagnetic
system.\cite{Delamotte2} This conjecture based on the non linear
$\sigma$ model ($d=2+\epsilon$ expansion) and on symmetry
arguments was checked by Southern and Young\cite{Southern_Young}
and Caffarel et al.\cite{Caffarel}

At higher temperatures the topological defects will clearly have a
role.\cite{Caffarel,Southern_Xu,Kawamura_O3_2d,Wintel_O3_2d}\\
Caffarel et al.\cite{Caffarel} have studied numerically two models
having identical spin waves but not the same topological defects.
One has the topological $Z_2$ defects, the other not. They showed
that the two models are equivalent at low temperatures, but show
differences at higher temperatures
due to the topological defects.\\
Southern and Xu\cite{Caffarel} studying by MC simulations the
vorticity associated to
the vortex--antivortex proposed that the vorticity has a jump at the critical
temperature similar to that of the KT transition.\\
Kawamura and Kikuchi\cite{Kawamura_O3_2d} using different boundary
conditions in MC simulations also observed various phenomena
associated to the vorticity
which, according to them, are a proof of a phase transition.\\
Last, Wintel et al.\cite{Wintel_O3_2d} have studied theoretically
and numerically the region above the suspected critical
temperature for the triangular antiferromagnetic lattice
($T_c\sim0.29$). They claimed that the correlation length and
susceptibility must follow a KT law.

There are numerical evidences of the importance of  topological
defects at finite temperatures and it seems that these systems
undergo a kind a topological phase transition. Nevertheless we
have no certainties that the phenomena present at finite sizes
would hold for the infinite size.

\subsection{Frustrated Heisenberg spin systems: $Z_2\otimes SO(3)$,
$Z_3\otimes SO(3)\,$\dots}

By inspecting Table~\ref{table_BS_O3} it is not difficult to see
that other order parameters exist in frustrated systems. One of
the most interesting should be $Z_2 \otimes SO(3)$. It corresponds
to a non planar GS which could exist in experimental systems.
Numerically the simplest system would be the Stiefel\index{Stiefel
model}  $V_{3,3}$ model or equivalently the Ising--$V_{2,2}$
model. The comparison with the $XY$ case would be very
interesting. In particular the Ising transition could appear near
the $SO(3)$ transition (see previous section) and the coupling
between the $Z_2$ walls and the vortex could be very instructive.
If we follow the second hypothesis of Olsson for the $XY$ case
(see section~\ref{2d_XY_Z2}), the transition should display an
``almost second-order transition''.

Symmetrically the coupling of the $SO(3)$ vortex with a $Z_3$
Potts model which appear in the Potts--$V_{3,2}$ model could also
disclose interesting properties. Does it show  first-order
properties as for $SO(2)$ vortex?

Some other breakdowns of symmetry could also appear like $Z_3
\otimes Z_2 \otimes SO(3)$ in the antiferromagnetic triangular
lattice with large NNN interaction. Similar questions appear. We
note that the $SO(3)/SO(2)$ order parameter does not have a
topological transition. The transition of the type $Z_2\otimes
SO(3)/SO(2)$ for the $J_1$--$J_2$ model or $Z_3\otimes
SO(3)/SO(2)$ for the triangular lattice with intermediate NNN
interaction should have a standard transition (Ising or
three-state Potts model).

\subsection{Topological defects for $N\geq 4$ }\index{topological defect}

Some other topological defects should exist for an order parameter
of the type $SO(N)$, with a GS in $N-1$ dimensions, $Z_2\otimes
SO(N)$, with a GS in $N$ dimensions, and $Z_q\otimes SO(N)$ with a
coupling of a GS in $N-1$ dimensions and a lattice symmetry
$\cdots$ All questions raised for Heisenberg spins still hold  in
these cases.

\section{General conclusions}

In this chapter we have studied the phase transition in frustrated
systems  between two and four dimensions. We have found various
breakdowns of symmetry, contrary to the unique one  for
ferromagnetic systems.

In three dimensions the transition is always of first order in the
thermodynamic limit.  However for ``small'' sizes in numerical
simulations or for temperatures not ``too close'' to the
transition temperatures in experiments, the system could display
an ``almost universality class'' for an $O(N)/O(N-2)$ breakdown of
symmetry. Many compounds studied experimentally are in this class.

In two dimensions the situation is much less clear. Indeed the
topological defects can play a fundamental role and their
couplings with a discrete symmetry (Ising or Potts) is unknown. We
hope that in the near future the two-dimensional case will be
clarified as the three-dimensional one on which our understanding
has increased considerably in the last decade.

\section*{Acknowledgments}
\addcontentsline{toc}{section}{Acknowledgments} I am grateful to
Prof. K. D. Schotte and Sonoe Sato for their support, discussions,
and for critical reading of the manuscript. This work was
supported by the DFG SCHO 158/10-1.


\section*{Appendix A: Monte\index{Monte Carlo simulation} Carlo
Simulation}\label{appendixMC}
\addcontentsline{toc}{section}{Appendix A: Monte Carlo Simulation
}


The purpose of this section is just to show the fastest algorithm
to run Monte Carlo simulations, the best method to distinguish a
first-order transition from a second-order one and to extract the
critical exponents with a reliable estimate of errors.

\vspace{0.4cm} {\bf Markov chains and algorithms}: Since we cannot
enumerate all the spin configurations, we are forced to use
numerical simulations to calculate the physical quantities. The
method of choice is the Monte Carlo method generating a set of
phase--space configurations from a Markov chain defined by the
transition probability $W[s,s']$ between states ${\lbrace s
\rbrace}$ and ${\lbrace s' \rbrace}$, where the new configuration
depends only on the preceding one. There are many ways to get the
transition probability $W[s,s']$.\cite{Creutz} The Metropolis
algorithm is the simplest but not very efficient one. A new
heat--bath algorithm, called Fast Linear
Algorithm,\cite{Loison_FLA} for vector spins is more efficient.
For example it is three times faster than the Metropolis algorithm
for a two-dimensional triangular antiferromagnetic lattice at the
critical temperature. Furthermore the use of the
over--relaxation\cite{Creutz} can reduce strongly the
autocorrelation time.\cite{Loison_TA_XY} The cluster algorithm
cannot be used for frustrated spin systems. Indeed there are two
problems. The first is to take into account the competition of
different interactions in a plaquette. It means that many spins
should be considered in one step when constructing the cluster.
Second, even if we were able to construct a cluster, the BS is no
more $O(N)/O(N-1)$, but $O(N)/O(N-2)$, for example. Consequently a
symmetry to a $N-1$ plane used in the Wolff's
algorithm\cite{Wolff89} should be changed.

Thermodynamic averages are taken simply by
\begin{equation}
\bar A ~=~ (1/N_{MC}) \sum_{t=1}^{N_{MC}} A[t]
\end{equation}
where $N_{MC}$ is the number of  new configurations generated
(Monte Carlo Steps) and $A[t]$ the value of the quantity at step
$t$. The great advantage of this method is that the partition
function $Z = \sum_{\lbrace s \rbrace} e^{- \beta E[s]}$ needs not
be calculated.

\vspace{0.4cm} {\bf To estimate the errors} of the averaged
quantities we have to take into account that each new
configuration is correlated with the previous one. We define the
autocorrelation time $\tau$ by the number of MC steps required to
obtain two uncorrelated spin configurations. It is calculated
using the autocorrelation function\cite{MullerB73} $C(t)=(1 /
\chi) [ \langle A(t)A(0) \rangle-\langle A \rangle^2 ]$ where time
is measured in MC steps (see comments in the appendix of Ref.
[22]). If $N \gg \tau$, then a useful approximation for the error
in $ \bar A$ is given by
\begin{eqnarray}
(\delta \bar A)^2 &=& \frac{\chi }{N_{MC}/(1+2 \tau)}\\
\label{susceptibility}
\chi &=& \langle A^2 \rangle -\langle A \rangle^2 \, .
\end{eqnarray}
This expression is identical to the standard deviation but with
an effective number of independent measurements given by
\begin{equation}
\frac{N_{MC}}{1+2 \tau}.
\end{equation}
Problems arise when quantities are a combination of different
averages, for example the susceptibility (\ref{susceptibility}).
We could try to treat $\langle A^2 \rangle$ and $\langle A
\rangle^2$ as independent quantities and estimate the error by the
sum of the errors of the two quantities. But the result will be
overestimated due to the correlation between the two elements of
the sum.  To solve this problem we can use, for example, the
Jackknife procedure.\cite{Jackknife} An application of this method
can be found in the appendix of Ref. [22].

\vspace{0.4cm} {\bf The histogram method}: The great advantage of
the histogram method in the analysis of MC data is that a run at a
single temperature $T_1$ can be used to extract results for a
continuous range of nearby temperatures.\cite{ferrs} In practice
the range of temperature over which $\langle A \rangle$ may be
estimated from a single MC run at $T_1$ is limited by the range
over which reliable statistics can be expected for $H(E)$, the
histogram of the energy. A rough guide is $T_a < T < T_b$, where
$T_a$ and $T_b$ correspond to the energies $\langle E_a \rangle$
and $\langle E_b \rangle$ at which $H(E) \simeq \frac12 H_{max}$.
Since $H(E)$ becomes more sharply peaked with system size, the
valid temperature range becomes smaller as $L$ increases. Multiple
histograms made at a number of nearby temperatures may be combined
to increase accuracy.

\vspace{0.4cm} {\bf Nature of the transition}: Differentiating a
weak first-order from a second-order transition could be
difficult. The finite size scaling (FSS) for a first-order
transition has been extensively studied
.\cite{Privman,Binder2,Billoire2} A first-order transition should
be identified by the following properties:
\begin{itemize}
\item[a)] The histogram of the energy, $P_T(E)$, has a double
peak. \item[b)] The minimum of the fourth order energy cumulant
$W$ varies as: $W=W^* + b\, L^{-d}$ where $W^*$ is different from
2/3. \item[c)] The temperatures $T(L)$ at which the specific heat
$C$ or the susceptibility $\chi$ has a maximum should vary as
$T(L) = T_c + a\, L^{-d}$. \item[d)] The maximum of $C$ and
$\chi$ are proportional to the volume $L^d$.
\end{itemize}
Obviously a double peak of $P_T(E)$ which becomes more pronounced
when the size increases is preferable (way a). If the two peaks
are too close, the three other possibilities (b, c and d) can
check if the probability is gaussian (second-order transition) or
not (first-order transition) in the limit of  infinite size.

\vspace{0.4cm} {\bf Second-order phase transition}: For a
second-order phase transition the interesting parameters are the
critical temperature and the critical exponents. There are at
least three main ways to calculate them:

\begin{itemize}

\item[a)] Consider the high and low temperature regions where the
correlation length $\xi \ll L$. There, we can fit $\xi=a_\xi
(T-T_c)^{-\nu}+{\rm corrections}$. The corrections become less
important near $T_c$, but then $\xi$ is very large, and a very big
system size is necessary. Furthermore the autocorrelation time
$\tau\sim(T-T_c)^{-z}$ becomes also very large and the error
$\delta\xi$ becomes very big even if we can use some tricks to
diminish it.\cite{LoisonPruessner}

\item[b)] The Finite Size Scaling (FSS) region is the most powerful method.
It is the region with
$\xi{\rm (theoretical)}\gg L$. The best method is to calculate the Binder
cumulant\cite{Binder81}
$U_L(T)~=~ 1~-\frac{\langle M^4 \rangle}{3\langle M^2 \rangle^2}$
for different size $L$.\\
{\bf 1.} We can calculate the critical exponents directly by
plotting $\chi L^{-\gamma/\nu}$ as a function of $U_L(T)$  and all
the curves must collapse for the correct exponent
$-\gamma/\nu=\eta-2$.\cite{Loison_O6_ferro} Indeed we can write
$\chi L^{\eta-2}=h(U_L)$ with $h$ an unknown function. The other
exponents $\beta/\nu$ and $\nu$ can be obtained similarly using
the magnetization and $V_1=\frac{\langle ME \rangle}{\langle M
\rangle}~-~\langle E \rangle$. It is the fastest way to get the
critical exponents. This method works well even for the
Kosterlitz-Thouless transition.\cite{LoisonO2Ferro}\\
{\bf 2.} We can also use $U_L(T)$ to calculate the critical
temperature $T_c$ using the crossing of $U_L$ for different sizes.
Very good statistics is needed because the evaluation of the
first correction is necessary
to get a correct value of $T_c$.\\
{\bf 3.} Having determined $T_c$ we can calculate the critical
exponents using, for example,
$\chi\propto L^{+\gamma/\nu}, \cdots$\\
{\bf 4.} The last properties are also valid for the maximum (or
minimum) of $\chi$ and $V_1$, but we need many simulations at
various temperatures to find the maxima because their locations
vary as a function of the size $L$ and are different for each
quantity. \item[c)] The dynamical
properties\cite{Janssen89,Huse89,Humayun91,Zheng98,Luo97} are not
very often used even if it is surely the fastest method available.
One has to prepare a state in the GS (i.e. $T=0$) or randomly
(i.e. $T=\infty$), and to observe the dynamical properties of the
system at the critical temperature $T=T_c$ for a finite number of
Monte Carlo steps before the equilibrium is reached. For an
example see Ref. [111].
\end{itemize}

\vspace{0.4cm} {\bf Kosterlitz--Thouless (KT)
transition}:\index{Kosterlitz-Thouless transition}  A KT
transition \cite{Thouless73} exists for two-dimensional lattices
with $XY$ spins. The unbounding of vortex--antivortex pairs appear
at the critical temperature, $T_c\sim0.9$ for square lattice.
\\
This transition has some special features: for $T<T_c$ the correlation
length ($\xi$) is infinite, while for $T>T_c$ it has an exponential decreasing
contrary to the power law behavior in the standard transition.\\
To find $T_c$ and the critical exponents we have several ways:
\begin{itemize}
\item[a)] we can fit $\xi$ in the high temperature region, which
is problematic because of the many free parameters and of the exponential form.
\item[b)] we can use a method using the behavior of various quantities in the
finite size scaling region (FSS) where $\xi\gg L$.\\
{\bf 1.}  The first one is to use the universal jump of the
helicity $\Upsilon$ at the critical temperature (in the finite
size scaling region where $\xi\gg L$), but this method requires
the jump of $\Upsilon(L)$ which is known for the non frustrated
case, but not for the frustrated case.
\\
{\bf 2.} It is therefore interesting to find another method to get
the critical exponent, even without calculating $T_c$. The method
(b.1.) introduced previously for a second-order phase transition
using the Binder parameter works well  for the KT
transition.\cite{LoisonO2Ferro} The Binder parameter, contrary to
the common belief, crossed around $T_c$. Plotting $\chi
L^{\eta-2}$ as a function of $U_L$ for various sizes, the curves
collapse at the correct value of $\eta=0.25$. This method can be
applied whatever the form of $\xi$ is. Therefore it should be
applicable to all phase transitions. \item[c)] The dynamical
behavior of this system can also be used .\cite{Zheng98}
\end{itemize}
For the two-dimensional ferromagnetic $XY$ spin system, all
methods mentioned above are in agreement.


\section*{Appendix B: Renormalization Group: Landau-Ginzburg theory,
expansions in fixed dimension $d=3$ and for
$d=4-\epsilon$ and its implications for
experiments}\index{renormalization group}\index{Landau--Ginzburg}
\label{appendixRG}
\addcontentsline{toc}{section}{Appendix B:
Renormalization Group}

In this short section  we would like to give the necessary knowledge to
help to understand the concepts of
fixed\index{fixed point} points, flow diagrams and crossovers.

First consider  a system of Heisenberg spins  with a  Hamiltonian:
\begin{eqnarray}
\label{equ_H}
H = -J_1 \sum_{(ij)} {\bf S}_{i}.{\bf S}_{j} \, ,
\end{eqnarray}
and the spin is restricted to be of norm $S=1$. Replacing this
constraint by an exponential potential $S=1 \Leftrightarrow
\int_0^\infty e^{u({\bf S}^2-1)^2}\cdot dS$ the Hamiltonian can be
written in the form
\begin{eqnarray}
\label{phi4}
H={K} \sum_{<ij>}
({\bf S}(i)-{\bf S}(j))^2
+r\sum_i {\bf S}(i)^2
+u_0 \sum_i \left({\bf S}(i)^2 \right)^2  \, .
\end{eqnarray}
Two remarks:\\
1.  Other terms could be added in the Hamiltonian. These
additional  terms could be unimportant that means they become
irrelevant near the fixed point (see later). Or we cannot treat
them, that means they are not renormalizable in technical terms.
If these neglected terms are important the method cannot describe
the physics of the phase transition. This unfortunate case occurs
for frustrated system for low dimensions
of space and spins.\cite{chapter_delamotte}\\
2. In addition, since the transformation is not exact, the
starting value  $u_0$ cannot be known even if it is possible to
make a rough guess.\cite{chapter_delamotte}

\begin{figure}[th]
\centerline{\psfig{file=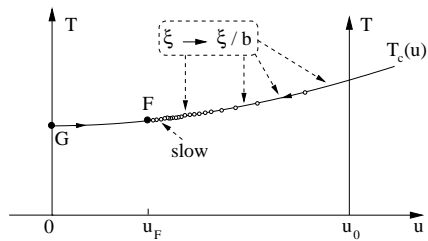,width=2.2in,angle=0}}
\vspace*{8pt}
\caption{
RG flow for (\protect{\ref{phi4}}). $F$ is the stable fixed point,
G is the Gaussian one. Near $F$ the flow is very slow.
\label{fig_RG_ferro}
}
\end{figure}

We have plotted in the Fig.~\ref{fig_RG_ferro} the RG flow for
this model.

The system will physically keep its value $u_0$. But its RG flow
will not. Following a series of steps the value of $u$ will change
in approaching the fixed point $F$. At each step the correlation
length of the system will be divided by a factor $b>1$. In
addition the flow becomes slower near the fixed point $F$.
Consequently, to reach the neighborhood of $F$, where the system
have a second-order transition, the initial $\xi$ must be large
``enough''. We know that $\xi(T)\sim (T-T_c)^{-\nu}$ for a
second-order phase transition. Hence, to obtain numerically or
experimentally $\xi$ which follows a power law (without too many
corrections), the temperature must be close ``enough'' to the
critical temperature $T_c$. Numerically, in the Finite Size
Scaling (FSS) regions where $\xi({\rm theoretical}$ is much bigger
than the size of the system $L$, the FSS law (see
\ref{appendixMC}) will be valid only for $L$ large ``enough''. We
observe that in the Fig.~\ref{fig_RG_tri} only the plane
$T_c(u,v)$ is plotted.

\begin{figure}[th]
\centerline{\psfig{file=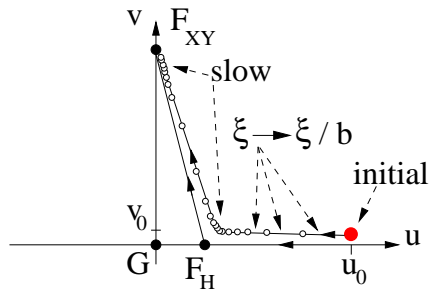,width=2.2in,angle=0}}
\vspace*{8pt} \caption{ Schematic RG flow ($T_c(u,v)$ plane) for
(\protect{\ref{phi4}} with $D\sum_i (S_i^z)^2$). $F_H$ is the
Heisenberg fixed point, $F_{XY}$ is the $XY$ fixed point, G is the
Gaussian one. Near $F_H$ and $F_{XY}$ the flow is very slow.
\label{fig_RG_ferro2} }
\end{figure}

Consider now a model with a $XY$ anisotropy, for example we add to
the Hamiltonian (\ref{equ_H}) a term $D\sum_i (S_i^z)^2$ with
$0<D\ll 1$. We have now two parameters in our Landau-Ginzburg
model: $u$ for the length of the spin and $v$ associated to $D$.
The critical plane $(u,v)$ of the flow diagram of this model is
plotted in Fig.~\ref{fig_RG_ferro2}. From the initial point
$(u_0,v_0\ll 1)$ the flow will go close to the Heisenberg fixed
point $F_H$ and then has a crossover to  the $F_{XY}$ fixed point.
Near $F_H$ the flow is very slow and needs a lot of steps to
escape from the influence of $F_H$ and reaches finally the
neighborhood of $F_{XY}$. Therefore, to observe the ``true'' $XY$
behavior, the correlation length must be very large.  That means
that the temperature must be very close of $T_c$ or, if we are in
the FSS region, the system size must be very large.

We must know the stability of the fixed point when anisotropies
are present. The ``quasi'' general rule is that the system goes to
the fixed point with the lowest spin symmetry and the biggest
space dimensions. For example consider the quasi one-dimensional
system CsMnBr$_3$. This compound will display a one-dimensional
behavior before reaching the three-dimensional behavior. Another
example is Cu(HCOO)$_2$2CO(ND$_2$)$_2$2D$_2$O which has a small
Ising anisotropy. In this case the system will show a Heisenberg
behavior before showing the Ising behavior. In real compounds
small ``anisotropies'' are almost always present. Therefore many
crossovers  will appear when the temperature approaches the
critical temperature. Consequently interpretations of experiments
could be difficult.

To get the critical exponent and the flow diagram, the expansion
in fixed dimension $d$ or in $d=4-\epsilon$ consists of expanding
the exponential $e^{Hamiltonian(u)}$ around the Gaussian fixed
point $u=0$. Results are in the form of a power series of $u$.
Unfortunately this series is not convergent, but in the
ferromagnetic case it has been shown to be resummable with
Pade--Borel techniques. However, there are many ways to resum and
one must be chosen following some criteria.\cite{LeGuillou} This
resummation has been proved efficient for ferromagnetic systems,
however it is not certain that it will work for frustrated
systems. Indeed the presence of two vectors $(O(N)/O(N-2)$ BS in
this case (compared to the $O(N)/O(N-1)$ BS in the ferromagnetic
case) gives rise to a series of power of $u$ and $v$ ($u$,$v$,
$uv$, $u^2$, $\cdots$) and this double expansion is difficult to
resum.

\end{document}